\documentclass[a4paper,twocolumn,11pt,accepted=2023-07-31]{quantumarticle}

\pdfoutput=1

\usepackage[table]{xcolor}
\usepackage[utf8]{inputenc}
\usepackage[english]{babel}
\usepackage{tikz}
\usepackage{graphicx}
\usepackage{hyperref}
\usepackage{lipsum}
\usepackage{color}
\usepackage{xcolor}
\usepackage{braket}
\usepackage[T1]{fontenc}
\usepackage{comment}
\usepackage{sidecap}
\usepackage[numbers,sort&compress]{natbib}
\usepackage{bbold}
\usepackage{amsmath}
\usepackage{amsthm}
\usepackage{amssymb}
\usepackage{braket}

\addto\captionsenglish{}

\newcommand{\lx}{\bar{{X}}}
\newcommand{\lz}{\bar{{Z}}}

\newtheorem{theorem}{Theorem}

\newtheorem{definition}[theorem]{Definition}

\begin{document}

\title{Correcting non-independent and non-identically distributed errors with surface codes}

\author{Konstantin Tiurev} \email{konstantin.tiurev@quantumsimulations.de}
\affiliation{HQS Quantum Simulations GmbH, Rintheimer Straße 23, 76131 Karlsruhe, Germany}
\author{Peter-Jan H. S. Derks}
\affiliation{Dahlem Center for Complex Quantum Systems, Freie Universit\"at Berlin, 14195 Berlin, Germany}
\author{Joschka Roffe}
\affiliation{Dahlem Center for Complex Quantum Systems, Freie Universit\"at Berlin, 14195 Berlin, Germany}
\author{Jens Eisert}
\affiliation{Dahlem Center for Complex Quantum Systems, Freie Universit\"at Berlin, 14195 Berlin, Germany}
\affiliation{Helmholtz-Zentrum Berlin f{\"u}r Materialien und Energie, 14109 Berlin, Germany}
\author{Jan-Michael Reiner}
\affiliation{HQS Quantum Simulations GmbH, Rintheimer Straße 23, 76131 Karlsruhe, Germany}

\maketitle

\begin{abstract}
A common approach to studying the performance of quantum error correcting codes is to assume independent and identically distributed single-qubit errors. However, the available experimental data shows that realistic errors in modern multi-qubit devices are typically neither independent nor identical across qubits. In this work, we develop and investigate the properties of topological surface codes adapted to a known noise structure by Clifford conjugations. We show that the surface code locally tailored to non-uniform single-qubit noise in conjunction with a scalable matching decoder yields an increase in error thresholds and exponential suppression of sub-threshold failure rates when compared to the standard surface code. Furthermore, we study the behaviour of the tailored surface code under local two-qubit noise and show the role that code degeneracy plays in correcting such noise. The proposed methods do not require additional overhead in terms of the number of qubits or gates and use a standard matching decoder, hence come at no extra cost compared to the standard surface-code error correction. 
\end{abstract}

\section{Introduction}

Instances of topological \emph{quantum error correcting codes}~(QECCs) have shown tremendous success in providing a guideline towards realizing fault-tolerant logical qubits~\cite{Kitaev-AnnPhys-2003,TopologicalQuantumMemory, PhysRevLett.108.180501,PhysRevA.86.032324, PhysRevLett.97.180501,LandahlColor,Kubica2018,Bombin_2015}. 
Compared to other, more sophisticated prescriptions they are comparably experimentally viable due to the local structure of stabilizers and low-weight syndrome measurement operations. The most famous among these is the surface code~\cite{Kitaev-AnnPhys-2003,TopologicalQuantumMemory,PhysRevLett.108.180501,PhysRevA.86.032324}, which only requires four-qubit parity measurements, admits fast and efficient decoding, and exhibits one of the highest thresholds among quantum codes with a two-dimensional qubit architecture. 

As other topological QECCs, the surface code turns a collection of noisy qubits into a more robust logical qubit by redundantly encoding information in a non-local way. Assuming that single-qubit errors occur independently with probability $p$, the error rate of the logical qubit can be made arbitrary small when $p$ is below a threshold value $p_{\textrm{th}}>0$~\cite{9781107002173,doi:10.1126/science.279.5349.342,Kitaev-AnnPhys-2003}. For a standard \emph{Calderbank-Shor-Steane}~(CSS) surface code in the presence of depolarizing noise and assuming error-free parity measurements, the error threshold is estimated to be around $p_{\textrm{th}} \approx 16$\% when a perfect-matching algorithm is used for decoding~\cite{BonillaAtaides2021,PhysRevLett.104.050504,Criger2018}. 

With the actual \emph{experimental implementation} of quantum error correcting codes moving closer to reality~\cite{Acharya2023,satzinger_realizing_2021,doi:10.1126/science.1253742,WallraffSurface,HoneywellFaultTolerant,Roadmap}, interest is shifting towards devising schemes for quantum error correction that would respect physical desiderata that arise from properties of the chosen hardware implementation. This reflects several aspects of the code design. In particular, the performance of an error correcting code heavily depends on the \emph{structure of noise} affecting the physical constituents. In order to accommodate this, steps have been taken recently to adapt quantum error correcting codes and decoding procedures to such kind of errors~\cite{arXiv:2201.07802,tiurev2023domain,BonillaAtaides2021,PhysRevLett.120.050505,higgott2022fragile,PhysRevLett.124.130501,Srivastava2022xyzhexagonal,Miguel2023cellularautomaton}. Given some classical knowledge of structures that errors commonly take, the error correction procedure can be tailored to more efficiently handle such structured noise, resulting in dramatically reduced logical error rates and increased thresholds. For instance, the noise threshold of the standard CSS surface code is noticeably reduced when physical qubits experience biased noise~\cite{BonillaAtaides2021}, e.g., qubits are more prone to dephasing than bit flips. Simply adapting the aspect ratio of the surface code allows to noticeably enhance the code performance~\cite{Lee2021}. References~\cite{arXiv:2201.07802,BonillaAtaides2021,PhysRevLett.120.050505,PhysRevLett.124.130501,higgott2022fragile} employ a more involved procedure, referred to as Clifford deformation, to tailor the code stabilizers and a decoder to a given noise structure. The two prominent examples are the XY~\cite{PhysRevLett.120.050505,PhysRevX.9.041031,PhysRevLett.124.130501} and the XZZX~\cite{BonillaAtaides2021,PRXQuantum.2.030345} codes, both significantly outperforming the CSS surface code for biased noise in terms of the threshold and sub-threshold scaling. Tailoring an error correction strategy to a known noise structure hence plays a crucial role for protecting quantum information, and is expected to be increasingly important as technological development progresses. 

Research within the field of QECCs typically assumes that all physical qubits that make up the code are identical, that is, every qubit experiences noise that can be modelled as an independent and identically distributed~(iid) process. In experimental reality, however, this assumption often turns out to be unrealistic. In modern quantum chips~\cite{ibmBrooklyn,ibmWashington,rigetiAspenm}, noise characteristics vary notably from one qubit to another~\cite{PhysRevA.106.062428, PhysRevA.108.022401, PhysRevLett.127.180501,harper2023learning}, which significantly compromises the performance of existing error correcting schemes~\cite{PhysRevA.106.062428,harper2023learning}. In certain devices, qubits not only have different error rates, but are also prone to different types of Pauli errors~\cite{PhysRevA.108.022401}, with Rigetti’s Aspen-M-2~\cite{rigetiAspenm,PhysRevA.108.022401} and IBM’s Washington~\cite{ibmWashington,PhysRevA.108.022401} being two such examples. Furthermore, experimental benchmarks show that errors can correlate across multiple qubits, either due to direct interactions~\cite{OGorman2016,PhysRevB.70.115310}, such as unwanted cross-talk~\cite{PhysRevLett.127.060505,PhysRevApplied.12.054023,harper2023learning,Xue2022,debroy2020logical}, or via coupling to a common bosonic bath~\cite{PhysRevA.89.042334}. Recent experiments have proven that such correlated errors have a crucial effect on the efficiency of surface-code quantum error correction~\cite{harper2023learning}. An optimal code design should respect a known non-iid noise structure of a particular device. However, despite the apparent practical importance, the performance and optimization of QECCs in the presence of non-iid errors have been barely investigated. Several works have discussed correction of multi-qubit correlated errors~\cite{Baireuther2018,PhysRevA.89.042334,PhysRevA.69.062313,PhysRevLett.96.050504,PhysRevA.89.032316,PhysRevA.90.042315}, while the research
of QECCs under non-identically distributed errors has been ignored until very recently and is limited to only a few publications~\cite{PhysRevA.106.062428, PhysRevA.108.022401, harper2023learning, martinez2022multiqubit,PhysRevX.9.021041}. 

In this work, we investigate the error correcting capabilities of the surface code in the presence of non-iid noise and propose ways to adapt the error correction procedure to more efficiently battle such noise. Firstly, we design a Clifford-deformed version of the surface code tailored to efficiently correct \emph{non-identically distributed Pauli errors} when used in conjunction with a scalable perfect-matching decoder~\cite{edmonds_1965}. We numerically benchmark the performance of the noise-tailored code versus non-optimized CSS surface codes and demonstrate that (i)~the sub-threshold logical error rates are exponentially suppressed as a function of the code distance and (ii)~the threshold grows monotonically as noise becomes less uniform across qubits. Secondly, we show that similar improvements in error correcting capabilities of the surface code can be achieved by modifying input parameters of a matching decoder in a situation where qubits experience \emph{non-identically distributed total error rates}. Finally, we study the behaviour of the surface code under a combination of single-qubit and local two-qubit errors. We show that the code stabilizers can be adapted to a known correlated noise structure to make syndrome configurations plausible for efficient perfect-matching decoding. Strikingly, our numerical simulations suggest that the sub-threshold scaling can be significantly improved even when no information about the noise structure is provided to the decoder. Instead, it suffices to merely account for code degeneracy~\cite{PhysRevLett.98.030501}, i.e., for a number of equivalent shortest-path logical operators, which is determined by a combination of a surface code geometry and a noise model. We expect this framework to be useful when tailoring codes to physical needs that arise in superconducting, trapped-ion, or photonic architectures.

The remainder of this article is organized as follows. Section~\ref{sec:surface_code} introduces the standard surface code and a decoder based on a \emph{minimum weight perfect matching}~(MWPM) algorithm~\cite{edmonds_1965,10.1063/1.1499754,Kolmogorov2009} and shows that the code can be locally transformed by a process known as Clifford deformation. Section~\ref{sec:nonuniform} introduces a variant of the surface code designed to correct non-identically distributed Pauli errors. In Sec.~\ref{sec:correlated}, we study the performance of the surface code under correlated errors between nearest-neighbor data qubits and point out a crucial role of code degeneracy for correcting such errors. We conclude in Sec.~\ref{sec:conclusions} by discussing potential directions for improving the performance of noise-tailored error correcting codes.

\section{Overview of the surface code}\label{sec:surface_code}

In order to appreciate the idea of tailoring the code design to noise of a known structure, we start by reviewing elements of the surface code. The basic idea of a QECC is to redundantly encode the state of a single logical qubit into a larger array of noisy physical qubits. Encoding is achieved by fixing $N-1$ independent parity check operators, or stabilizers, on $N$ bare qubits, with the remaining degrees of freedom corresponding to the logical qubit. Throughout the course of computation, errors are detected by measuring stabilizers and subsequently corrected. If bare qubit errors are sufficiently rare, i.e., occur with probability below the threshold value $p_{\textrm{th}}$, the logical error rate can be made arbitrary small by increasing a number of physical qubits in the code. The challenge is to design practical QECCs that allow to detect and correct the most common errors in realistic experimental settings. Among other choices, the surface code offers particularly high thresholds while requiring only parity checks within sets of nearest-neighbor qubits in a two-dimensional qubit layout. Below we describe the basic operation of the surface code and the decoding procedure. 

\subsection{CSS surface code}

\begin{figure}
\centering
  \includegraphics[width=0.9\columnwidth]{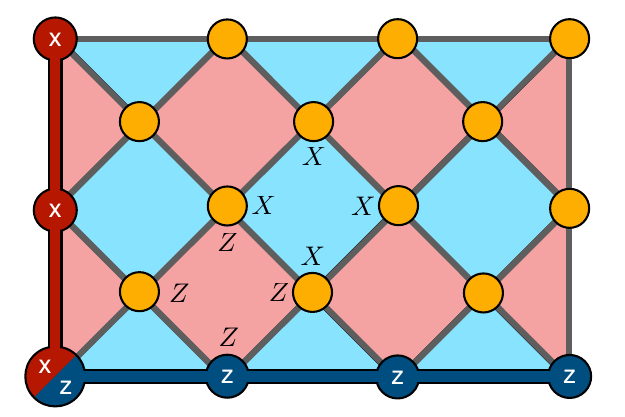}
  \caption{\label{fig:1} The CSS surface code. Physical qubits are arranged on the edges on a rectangular lattice and shown with orange circles. Star and plaquette stabilizers are defined in Eq.~\eqref{eq:CSS_stabilizers} and shown with blue and salmon tiles, respectively. Logical $\bar{X}$~($\bar{Z}$) operators are chains of single-qubit Pauli operators shown in red~(blue).}
\end{figure}

The surface code is a stabilizer quantum error correcting code formed of qubits placed on the edges of a 2D square lattice, as shown in Fig.~\ref{fig:1}. Particularly, the surface code of $d_1\times d_2$ qubits arranged on lattice $L$ contains $N = d_1d_2 + (d_1-1)(d_2-1)$ qubits and has a distance $d=\textrm{min}(d_1,d_2)$. Since the original surface code is a CSS code, it is referred to as the CSS surface code, following the
nomenclature of Ref.~\cite{arXiv:2201.07802}. 
The code space of the surface code is described by its stabilizer group $\mathcal{S}$, which is an Abelian subgroup of the $N$-qubit \emph{Pauli group} $\mathcal{P}_N$,
the group generated by products of single-qubit Pauli operators 
together with suitable scalar factors. Particularly, the stabilizer group of the standard surface code is generated by two types of stabilizers, namely, \emph{star} and \emph{plaquette} operators defined as
\begin{equation}\label{eq:CSS_stabilizers}
A_{\Diamond} := \prod_{i \in \Diamond} X_i
\quad
\textrm{and}
\quad
B_{\square} := \prod_{j \in \square} Z_j,
\end{equation}
respectively. Here and in the following, 
$X_i$, $Y_i$ and $Z_i$ denote the respective qubit Pauli operators applied to qubit with vertex label $i\in L$.
We will also use $X,Y,Z$ for the abstract Pauli
operators when the support is not specified.
$\Diamond$ and $\square$ denote cells as subsets of $L$ of four nearest-neighbor qubits and form, respectively, primal and dual sub-lattices illustrated in Fig.~\ref{fig:1}. Operators on the logical qubit are defined as chains of single-qubit Pauli operators connecting opposite edges of the lattice and commuting with all stabilizers. Measuring stabilizers hence does not reveal information about the state of the logical qubit. With the above definition~\eqref{eq:CSS_stabilizers}, a logical $\lx$~($\lz$) operator corresponds to any chain of single-qubit $X$~($Z$) operators connecting top and bottom~(left and right) edges of the lattice in Fig.~\ref{fig:1}. Basis state vectors of the logical qubit are then defined in a standard way,
\begin{align}\label{eq:logical_states}
\lz \ket{0}_L &= (+1)\ket{0}_L,
&
\lz \ket{1}_L &= (-1)\ket{1}_L
\end{align}
and, by construction, obey
\begin{align}\label{eq:logical_operators}
\lx \ket{0}_L &= \ket{1}_L,
&
\lx \ket{1}_L &= \ket{0}_L.
\end{align}

Stabilizers~\eqref{eq:CSS_stabilizers} are measured throughout the computation and the result of their measurement, also called the error syndrome, indicates which error has occurred. We note that stabilizer measurement is itself a process prone to errors and the measurement apparatus can in principle return incorrect eigenvalues. In such a case, repetitive extraction of syndromes is necessary to define the correct syndrome configuration. Here we restrict our analysis to a simpler case of perfect syndrome measurements and only consider errors that occur in data qubits. The code tailoring techniques introduced in this paper only require single-qubit Clifford rotations on data qubits. Since no additional multi-qubit operations are involved, such code tailoring does not introduce additional error propagation paths between the qubits of the code. Hence, the gain in performance observed for perfect stabilizer measurement will preserve on the fault-tolerant regime too.

For the CSS surface code defined according to Eq.~\eqref{eq:CSS_stabilizers}, a single-qubit $X$~($Z$) error commutes with all star~(plaquette) stabilizers and therefore only creates pairs of defects, or anyons, in the dual~(primal) sub-lattice. Decoding is performed by finding the most probable chains of errors corresponding to a measured syndrome. The most commonly used decoder is based on the MWPM algorithm. It first constructs an input graph by assigning each measured defect a vertex, with weighted edges corresponding to chains of errors connecting two defects. Assuming that single-qubit $X$ and $Z$ errors occur with equal probability $p$, weights $w_{i,j}\geq 0$ between vertices $i,j$ are chosen according to
\begin{equation}\label{eq:weight_css}
    w_{i,j} := -\log P_{i,j},
\end{equation}
where
\begin{equation}\label{eq:prob_css}
    P_{i,j} := p^{l_x+l_z}
\end{equation}
is the probability of connecting two stabilizers that have horizontal distance $l_x$ and vertical distance $l_z$ by the shortest chain of single-qubit errors. Substituting Eq.~\eqref{eq:prob_css} into Eq.~\eqref{eq:weight_css} and ignoring a common coefficient $-\log(p)>0$, the edges of the input graph are weighted according to the \emph{Manhattan distance}
\begin{equation}\label{eq:weight_manhattan}
    w_{i,j} = l_x + l_z,
\end{equation}
which is the distance induced by the vector $l_1$-norm.

The algorithm then searches for a perfect matching such that the sum of the weights of the edges is minimal, corresponding to a maximally probable error chain, and returns indices of defects that should be locally paired by the correction. MWPM in its standard implementation assumes single-qubit $X$ and $Z$ errors independently, hence decoding takes place separately in dual and primal sub-lattices. Pauli $Y$ errors create pairs of defects in the two sub-lattices simultaneously, which compromises the capabilities of the standard matching decoder. Certain modified versions of the MWPM algorithm can, to some extent, take correlations beyond the paradigm of independent $X$ and $Z$ errors into account and therefore demonstrate the improved decoding efficiency in the presence of $Y$ errors~\cite{PhysRevLett.124.130501,6874997}. For simplicity, here we will focus on the matching algorithm in its standard form. However, we note that Clifford deformation methods suggested in this paper can in principle be combined with more advanced versions of the MWPM decoder, e.g. those taking into account correlations between deformed sub-lattices~\cite{6874997} or allowing for parallel decoding~\cite{skoric2023parallel}.

Finally, there are classes of decoders not based on matching algorithms. As such the tensor-network decoder of Ref.~\cite{PhysRevA.90.032326} is capable of finding the optimal code-capacity thresholds, however, for the price of runtime scaling exponentially with the code size, rendering it incompatible with fast on-the-go error correction in real quantum hardware. 

\subsection{Clifford-deformed surface codes}\label{sec:deformation}

In the previous Section, we have constructed a CSS surface code according to the stabilizers of Eq.~\eqref{eq:CSS_stabilizers}. Equivalently, one can choose an alternative set of stabilizers by applying single-qubit Clifford transformations to the stabilizers of the CSS code. Building on the steps taken in Ref.~\cite{arXiv:2201.07802} we refer to any code derived in such a way as a Clifford-deformed QECC. 

In a Clifford-deformed surface code, the quantum states of each of the physical qubits $i\in L$ of the code are conjugated with a unitary $U\in C_1/U(1)$ from the single-qubit \emph{Clifford group} of order 24. The Clifford group is the group of unitaries that maps Pauli operators onto Pauli operators under conjugation. Specifically, the Hadamard  gate $H$ maps Pauli $X$ to Pauli $Z$ operators, while $H_{YZ} = H\sqrt{Z}H$ (both contained in the Clifford group) intertwines between Pauli $Y$ and $Z$. In what follows, the term Clifford deformation will refer to a replacement of a qubit Pauli operator by a different Pauli operator according to one of those conjugations~\cite{note_coherent_noise}. 

Any set of operators derived via Clifford de\-for\-ma\-tions---in the form of Clifford con\-ju\-ga\-tions---will form legitimate code stabilizers, since such conjugations preserve commutativity. Indeed, applying a unitary $U_i\in C_1/U(1)$ to Pauli operators $X$ and $Z$ measured on qubit $i$ transforms the operators as
\begin{equation}
\label{eq:transform}
    \begin{aligned}
        X &\mapsto U_i X U_i^{\dagger},
        \\
        Z &\mapsto U_i Z U_i^{\dagger},
    \end{aligned}
\end{equation}
and preserves anti-commutation,
\begin{equation}
\label{eq:anticomm}
    \{U_iXU_i^{\dagger},U_iZU_i^{\dagger}\} 
    = 
    \{X,Z\} = 0.
\end{equation}
Since operators measured on other qubits are intact, commutation of stabilizers is preserved and a unitary $U_i$ can be chosen for each qubit $i$ independently. 

Therefore, the surface code admits local Clifford deformations that can be chosen independently for each physical qubit. In the focus of this work are new instances of Clifford-deformed surface codes, following up on and building upon Ref.~\cite{PhysRevLett.120.050505}. We start by discussing examples of uniformly deformed surface codes in Sec.~\ref{subsec:biased} and subsequently introduce noise-tailored local deformations in Sec.~\ref{sec:nonuniform}.

\subsection{Surface codes for biased noise}\label{subsec:biased}

Clifford-deformed surface codes have demonstrated an exceptional performance when physical qubits are affected by biased noise~\cite{BonillaAtaides2021,PhysRevLett.120.050505}. Consider an error model in which each qubit is independently subject to a diagonal \emph{Pauli channel} $\mathcal{E}$ of the form
\begin{equation} \label{eq:pauli_channel}
    \mathcal{E}(\rho)
    :=
    (1-p)\mathbb{1}
    +
    p_x 
    X\rho X
    +
    p_y
    Y\rho Y
    +
    p_z
    Z\rho Z,
\end{equation}
where $p_x,p_y,p_z \in [0,1]$ 
are the probabilities 
of the corresponding single-qubit Pauli error and $
p=p_x + p_y + p_z \in [0,1]$ 
is a physical qubit error rate. This being a random unitary process, it is clear that $\mathcal{E}$ is a valid quantum channel for all allowed parameters. In particular, noise biased towards dephasing corresponds to 
$p_z> p_x, p_y$.

\begin{figure}[t]
    \centering
    \includegraphics[width=0.95\columnwidth]{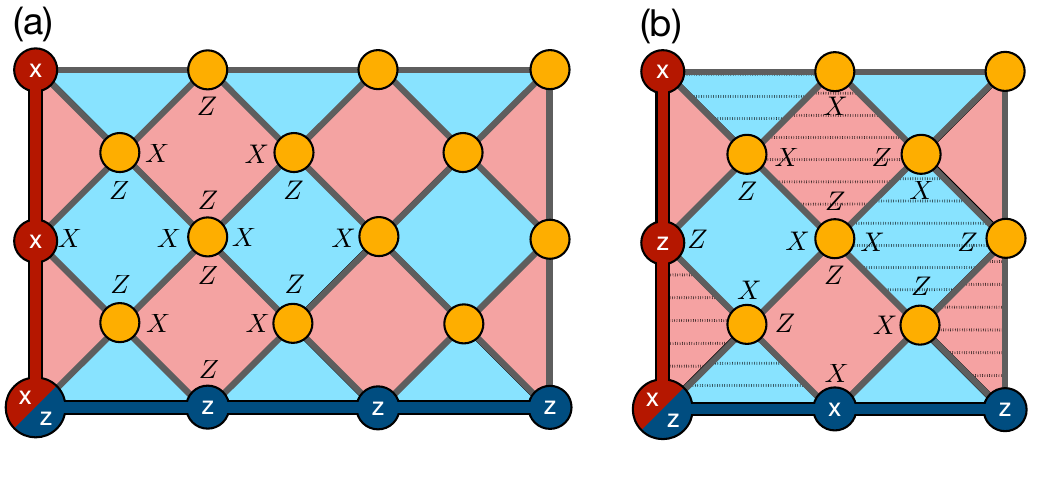}
    \caption{\label{fig:2} Surface codes tailored to biased noise in the (a)~XZZX and (b)~XXZZ configurations. (a)~In the XZZX code, each code stabilizer is the product of two Pauli $X$~(on horizontally located qubits) and two Pauli $Z$~(on vertically located qubits) operators. Logical $\bar{X}$~($\bar{Z}$) operator is shown with solid red~(blue) line and corresponds to a product of single-qubit $X$~($Z$) Pauli operators. (b)~In the XXZZ code the stabilizers are of two types. Stabilizers on even diagonals are products of Pauli $X$ operators on qubits to the north and west and Pauli $Z$ operators on qubits to the south and east, as shown with plain tiles. Stabilizers on odd diagonals differ by interchanging $X$ and $Z$ Paulis and are shown with shaded tiles. Logical $\bar{X}$ and $\bar{Z}$ operators are products of alternating $X$ and $Z$ single-qubit Pauli operators.}
\end{figure}

\begin{figure}[t]
    \centering
    \includegraphics[width=0.95\columnwidth]{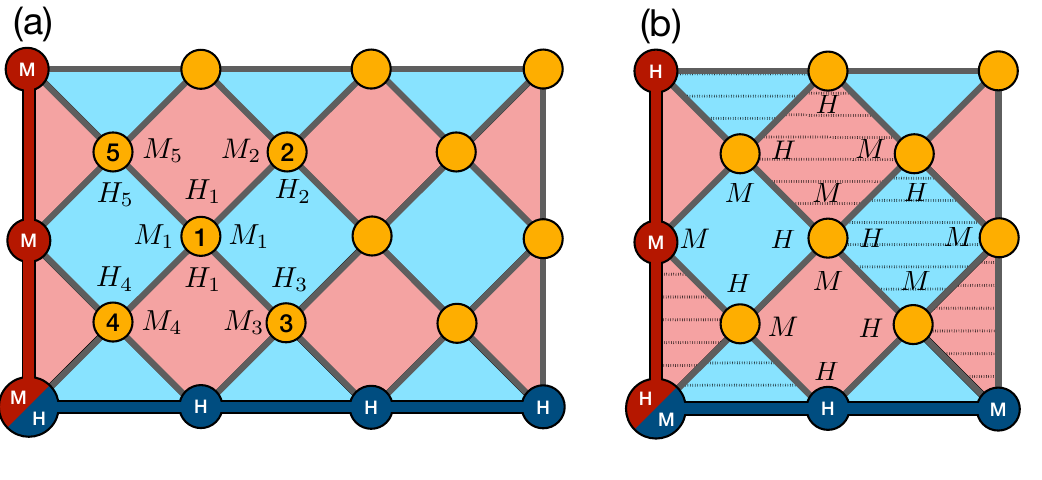}
    \caption{\label{fig:3} Noise-tailored surface codes in the (a)~MHHM and (b)~MMHH configurations. On each given qubit, stabilizers measure Pauli operators corresponding to medium and high rate errors. (a)~Operators $H_i$ and $M_i$ correspond to high-rate and medium-rate Pauli errors of qubit $i\in L$ they are measured on. Stabilizer $s$ of the MHHM code measures operators $M_i$~($H_i$) on qubits
    $i$ in the neighborhood of $s$ located to the horizontally~(vertically) from it. Logical $\lx$ and $\lz$ operators are products of Pauli operators corresponding to medium and high rate local errors, respectively. (b)~As in (a), operators $M_i$ and $H_i$ are chosen locally accordingly to error rates of a given qubit. Stabilizers of the MMHH code located on even diagonals~(solid tiles) differ from the stabilizers located on odd diagonals~(dashed tiles) by interchanging operators $M_i$ and $H_i$. For brevity, subscripts of single-qubit operators composing stabilizers of (b) and logical operators in (a,b) are omitted in the figure.}
\end{figure}

Reference~\cite{PhysRevLett.120.050505} has used a modified surface code, where the transformation $H_{YZ}$ has been applied to all qubits of the code. Such transformation leaves Pauli $X$ unchanged, while transforming the plaquette stabilizers to
\begin{equation}\label{eq:XY_stabilizers}
    B_{\square} = \prod_{j\in \square}Y_j.
\end{equation}
The resulting code is therefore referred to as the
\emph{XY code}. Alternatively, the \emph{XZZX code}~\cite{BonillaAtaides2021} can be obtained from the CSS code by applying a Hadamard transformation to qubits on even rows of the code. This effectively replaces both the star and plaquette stabilizers by
\begin{equation}\label{eq:XZZX_stabilizers}
    A_{\Diamond} = B_{\square} = X_{\textrm{w}}Z_{\textrm{s}}Z_{\textrm{n}}X_{\textrm{e}},
\end{equation}
where single-qubit Pauli operators acting on north, east, west, and south qubits within each stabilizer, 
and given the respective indices, 
are oriented as shown in Fig.~\ref{fig:2}~(a). 

The improved performance of the deformed codes originates from the amount of information available to the decoder. As an illustration, assume an infinitely biased noise described by a \emph{phase-flip channel}
\begin{equation}\label{eq:Z_channel}
\mathcal{E}'(\rho)
:=
(1-p)\mathbb{1}
+
p Z\rho Z.
\end{equation}
Since Pauli $Z$ operators commute with all stabilizers $B_{\square}$ of the form~\eqref{eq:CSS_stabilizers}, the dual sub-lattice of the CSS code never detects such error events and hence does not provide any useful information to the decoder. On the other hand, transforming dual stabilizers according to Eq.~\eqref{eq:XY_stabilizers} doubles the amount of useful syndrome information available to the decoder in the XY code. Similarly, Pauli $Z$ errors always create defects in either dual or primal sub-lattice of the XZZX code, depending on which physical qubit has been affected by an error event. 

The important difference between the two codes is in the decoder that can be employed to find the correct error configuration. In the case of the XY code, a single-qubit $Z$ error event simultaneously creates two pairs of anyons in both sub-lattices. When decoding such syndrome configurations, decoders based on perfect matching~\cite{PhysRevLett.124.130501} behave suboptimally, even when correlations are partially taken into account~\cite{6874997}. 
Hence, although both the primal and dual sub-lattices of the code contain information about the occurred errors, decoding such configurations require alternative, usually much more computationally demanding methods, such as the maximum likelihood decoding~\cite{PhysRevA.90.032326,PhysRevX.9.041031}. On the other hand, the same single-qubit $Z$ error only creates anyons in one of the sub-lattices of the XZZX code at a time, making it well-suitable for perfect-matching decoding. Below we will aim at constructing codes that admit efficient decoding with MWPM algorithms as they, due to their fast performance, can potentially be implemented for practical error correction in quantum hardware.

\subsection{Tailoring aspect ratios}

The XZZX code as in Fig.~\ref{fig:2}~(a), i.e., defined on a non-rotated lattice, performs non-optimally on a square-shaped lattice due to the structure of logical operators. For a square surface code of distance $d$, the probabilities of logical $\lx$ and $\lz$ errors are 
$\textrm{P}^{(\textrm{sq})}(\lx) = \mathcal{O}(p_x^d)$ 
and 
$\textrm{P}^{(\textrm{sq})}(\lz) = \mathcal{O}(p_z^d)$, 
respectively. Hence, for biased noise with
\begin{equation}
\eta := p_z/p_x \gg 1,
\end{equation}
the logical errors of a square-shaped XZZX code occur predominantly due to the logical $\lz$ error,
\begin{equation}
    \textrm{P}^{(\textrm{sq})}_{\textrm{fail}}
    \approx
    \textrm{P}^{(\textrm{sq})}(\lz) 
    \gg
    \textrm{P}^{(\textrm{sq})}(\lx).
\end{equation}

The failure rate of the code can be minimized by arranging the qubits on a rectangular lattice with an aspect ratio $A = d_1/d_2= \mathcal{O}[\log(p_z/p_x)]$, 
as in Fig.~\ref{fig:2}~(a). For a rectangular surface code containing the same number of qubits as a distance-$d$ squared code,
\begin{equation}
    d^2 + (d-1)^2 = d_1d_2 + (d_1-1)(d_2-1),
\end{equation}
the logical failure rate obeys 
\begin{equation}
    \textrm{P}^{(\textrm{r})}_{\textrm{fail}}
    \approx
    \textrm{P}^{(\textrm{r})}(\lz) 
    \approx
    \textrm{P}^{(\textrm{r})}(\lx)
    \ll
    \textrm{P}^{(\textrm{sq})}_{\textrm{fail}},
\end{equation}
leading to greatly suppressed logical failure rates in the regime of a large bias~\cite{Lee2021}. Furthermore, all paths connecting two anyons labelled by $i$ and $j$ in the XZZX code are realized with an equal probability
\begin{equation}\label{eq:prob_xzzx}
P_{i,j} = p_x^{l_x}p_z^{l_z}.
\end{equation}
Weights of the input matching decoder graph~\eqref{eq:weight_css} hence yield a weighted Manhattan distance,
\begin{equation}\label{eq:weighted_Manhattan}
w_{i,j} = -\log(p_x) l_x - \log(p_z) l_z.
\end{equation} 

For completeness, 
we introduce an alternative configuration of the surface code that yields optimal logical error rates on a square lattice for arbitrary noise bias. We refer to this code as the XXZZ code owing to the stabilizers configuration shown in Fig.~\ref{fig:2}~(b). For such a code, logical operators $\lx$ and $\lz$ are products of alternating per-qubit Pauli $X$ and $Z$ operators. For sufficiently large lattices, the probabilities of logical errors $\lx$ and $\lz$ are, therefore, approximately equal independently of the noise bias,
\begin{equation}\label{eq:equal_operators}
\textrm{P}^{\textrm{(sq)}}(\lx) 
\approx 
\textrm{P}^{\textrm{(sq)}}(\lz) 
\approx 
(p_x p_z)^{d/2}.
\end{equation}
Therefore, the aspect ratio $A=1$ can be kept constant for any bias~\cite{note_rotated_code}. 

\section{Non-identically distributed errors}\label{sec:nonuniform}

Physical qubits that constitute a QECC are conventionally assumed to be identical. However, available measurement data shows that noise characteristics in modern multi-qubit devices can vary significantly from one qubit to another, with some qubits being noisier that others~\cite{PhysRevA.106.062428, PhysRevA.108.022401, PhysRevLett.127.180501,harper2023learning,10.1145/3297858.3304007,10.1145/3510857,doi:10.1038/s41586-019-1666-5,https://doi.org/10.48550/arxiv.2010.07965,martinez2022multiqubit}. In addition, different qubits within one device can be prone to different types of Pauli errors, e.g., such that some qubits are more susceptible to phase flips, while others are more prone to bit flips~\cite{PhysRevA.108.022401}. The probabilities of different Pauli error channels can be expressed using qubit coherence times $T_1$~(relaxation time) and $T_2$~(dephasing time) as
\begin{equation}
    \begin{aligned}
        p_x &= p_y = \frac{1}{4}
        \Big{(} 
        1
        -
        {\rm e}^{-\frac{t}{T_1}}
        \Big{)},
        \\
        p_z &= \frac{1}{4}
        \Big{(} 
        1
        +
        {\rm e}^{-\frac{t}{T_1}}
        -
        2{\rm e}^{-\frac{t}{T_2}}
        \Big{)}.
    \end{aligned}
\end{equation}
Hence, qubits limited by their $T_1$ are more prone to $X$ and $Y$ errors, while qubits limited by $T_2$ are more prone to $Z$ errors. Experimental data shows that coherence times $T_1$ and $T_2$ depend heavily on hardware platform, design and manufacturing parameters of a particular device. As such, the parameter $T_1/T_2$ is relatively uniform across qubits in superconducting chips from Google Quantum AI and Zuchongzi~\cite{PhysRevLett.127.180501}. On the other hand, in devices such as Rigetti’s Aspen-M-2~\cite{rigetiAspenm} and IBM’s Washington~\cite{ibmWashington}, coherence times of different qubits within the same chip are vastly different, as shown by the calibration data presented in Ref.~\cite{PhysRevA.108.022401}. The noise model that takes into account non-uniformities in total and relative Pauli error rates captures the differences in the noise parameters of real qubits.

In this Section, we assume that noise channels of individual qubits can be measured. In Subsections~\ref{sec:mhhm_deformation}--\ref{sec:realistic_model}, we construct and benchmark Clifford-deformed surface codes locally tailored to efficiently correct non-uniform Pauli noise. In Subsection~\ref{sec:tailored_decoder}, we show that non-uniform total noise can be more efficiently corrected when the MWPM decoder is provided with information about local error rates of individual qubits. In Subsection~\ref{sec:inid}, we combine the two methods for correcting an arbitrary non-identically distributed noise. 

\subsection{Noise-aware Clifford deformations}\label{sec:mhhm_deformation}

Our aim is to consider Clifford deformations such that the most probable syndrome configurations for a given noise model are maximally compatible with perfect-matching decoding. As discussed in the previous section, the efficiency of such an MWPM decoder is compromised by interactions between the dual and primal lattices, i.e., by single-qubit Pauli error events that simultaneously create defects in all four stabilizers surrounding the qubit. One can minimize such events by applying
noise-tailored local Clifford deformations introduced earlier in Sec.~\ref{sec:deformation}. \smallskip\smallskip

\begin{definition}
[Clifford deformed codes for
non-uni\-form Pauli diagonal noise]
{\it Consider a qubit located at site $i\in L$ and a set of stabilizers $S_i$ which have that qubit in their support. The qubit is subject to Pauli diagonal noise channels with weights $p_x^{(i)},p_y^{(i)},p_z^{(i)}\in [0,1]$. Define $H_i\in\{X_i,Y_i,Z_i\} $ as the Pauli operator reflecting the maximum weight $h_i:=\max\{p_x^{(i)},p_y^{(i)},p_z^{(i)}\}$ in the Pauli channel and $M_i$ as the Pauli operator corresponding to medium weight $m_i := \max\{p_x^{(i)},p_y^{(i)},p_z^{(i)}\}\backslash \{h_i\}$. In the MHHM code, stabilizers from $S_i$ located east and west (north and south) of qubit $i$ measure operators $M_i$~($H_i$) on that qubit, as illustrated in Fig.~\ref{fig:3}~(a). The MMHH code is similar, but with stabilizers chosen as in Fig.~\ref{fig:3}~(b).}
\end{definition}
As an example, consider qubit $i$ with $p^{(i)}_x<p^{(i)}_z<p^{(i)}_y$. Then, in the MHHM code the stabilizers located horizontally and vertically from qubit $i$ measure Pauli $Y$ and $Z$ operators on that qubit, respectively. The commutativity of stabilizers is guaranteed by the anti-commutativity of $Y$ and $Z$, as in Eq.~\ref{eq:anticomm}. Since for each qubit the transformation~\eqref{eq:transform} can be chosen independently, the MHHM and the MMHH codes can always be constructed according to known error rates of qubits in the code. We refer to two possible layouts of a code locally optimized for perfect-matching decoding as the MHHM~[Fig.~\ref{fig:3}~(a)] and MMHH~[Fig.~\ref{fig:3}~(b)] surface codes due to their similarity to the XZZX~[Fig.~\ref{fig:2}~(a)] and XXZZ~[Fig.~\ref{fig:2}~(b)] codes, respectively. 

\subsection{A paradigmatic toy example}\label{sec:toy_example}

The exact model of how the noise bias varies from one qubit to another depends heavily on a particular hardware implementation of physical qubits. Our aim here is to provide a convincing argument that tailoring the code to local noise structure can be advantageous for decoding. We first demonstrate the proposed scheme on a simple toy model. Assume that the probabilities of low~($l$), medium~($m$), and high-rate~($h$) errors are identical across qubits, but for each qubit located at site $i\in L$ we choose
\begin{equation}\label{eq:permutation}
\{p_x^{(i)}, p_y^{(i)}, p_z^{(i)}\}
=
\pi_i\Big{(}\{m, l, h \} \Big{)},
\end{equation}
where $\pi_i$ denotes a random permutation applied at a qubit $i$, i.e., for each qubit we randomly assign low, medium, and high rates to different Pauli errors. This is not meant to be a realistic noise model,
but has been designed to explain the functioning of the approach. 
Using the known rates of Pauli errors at each individual qubit, one can transform the CSS surface code into the MMHH or MHHM code by applying local Clifford deformations to it. It is easy to see that the Clifford-deformed MMHH~(MHHM) code in the presence of the noise described by Eq.~\eqref{eq:permutation} is equivalent to the XXZZ~(XZZX) code affected by the noise with $\pi_i = 1$. Put another way, noise-informed Clifford deformations effectively undo the permutations applied to errors in Eq.~\eqref{eq:permutation}, turning the surface code into one of the biased-tailored codes.

\subsection{A more realistic noise model}\label{sec:realistic_model}

We will now investigate the code performance in a more realistic error model, where the total error probabilities as well as the relative probabilities of different Pauli errors vary between qubits. For a qubit located at site $i \in L$ we generate the error probabilities in two steps. First, we choose the error probabilities according to
\begin{equation}\label{eq:probabilities}
    p_k^{(i)} \sim \mathcal{N}(0.5, \sigma_{\textrm{P}}),
\end{equation}
where $k = \{x,y,z\}$ and $\mathcal{N}(\mu, \sigma_{\textrm{P}})$ is a truncated at $0$ and $1$ normal distribution with mean $\mu$ and standard deviation $\sigma_{\textrm{P}}$. Next, we normalize Pauli error probabilities $p_k^{(i)}$ on local error rates $p^{(i)}$ of individual qubits,
\begin{equation}\label{eq:constraint_var}
    \sum_{k=X,Y,Z}p_k^{(i)} = p^{(i)},
\end{equation}
where $p^{(i)}$ is also generated according to the normal distribution
\begin{equation}\label{eq:probabilities_var}
    p^{(i)} \sim \mathcal{N}(p, p\sigma_{\textrm{tot}})
\end{equation}
for each site $i\in L$. This process generates the probabilities of low, medium, and high rate errors independently for each qubit and randomly assigns these probabilities to Pauli errors. Noise in our model is hence characterized by an average error rate $p$ and a pair of parameters $\sigma = (\sigma_{\textrm{P}}$, $\sigma_{\textrm{tot}})$ defining, respectively, variations in types of Pauli errors and in total error rates of qubits. We will refer to the noise model described by $\sigma = (\sigma_{\textrm{P}}>0,0)$ as \emph{uniform total noise}, and to the noise model with $\sigma = (0,\sigma_{\textrm{tot}}>0)$ as \emph{uniform Pauli noise}. A trivial case of $\sigma = (0,0)$ corresponds to the iid error model. Knowing the rates of Pauli errors of each qubit, the noise-tailored surface codes can now be constructed according to the patterns of Fig.~\ref{fig:3}, as in the previous toy example. 

\begin{table}[t]
    \centering
    \begin{tabular}{|lcccc|}
    \cline{1-5} 
    $\sigma_{\textrm{tot}}$ & 0
    & 0
    & 0.5
    & 0.5
    \\
     $\sigma_{\textrm{P}}$ & 0
    & 0.5
    & 0
    & 0.5
    \\
    \hline
    \multicolumn{5}{|c|}{\cellcolor{gray!25} Manhattan distance metrics}
    \\
    \hline
    CSS  & 0.162 & 0.163 & 0.166 & 0.164
    \\
    MMHH & 0.164 & 0.185 & 0.165 & 0.185
    \\
    MHHM & 0.164 & 0.187 & 0.164 & 0.188
    \\
    \hline
    \multicolumn{5}{|c|}{\cellcolor{gray!25} Dijkstra distance metrics}
    \\
    \hline
    CSS  & 0.163 & 0.168 & 0.178 & 0.181 
    \\
    MMHH & 0.162 & 0.190 & 0.177 & 0.205
    \\
    MHHM & 0.165 & 0.190 & 0.178 & 0.206
    \\
    \hline
    \end{tabular}
\caption{Error correction thresholds of different surface codes determined in Appendix~\ref{sec:thresholds}. The first column provides the thresholds observed for a uniform total error model described by $\sigma=(1/2,0)$. The second column corresponds to a non-uniform error rate with $\sigma=(0,1/2)$. In the third column, both total and Pauli error rates vary between qubits. We benchmark the performance of decoders with weights determined according to two distance metrics. The first metrics is defined according to the Manhattan distances of Eq.~\eqref{eq:weight_manhattan}~[CSS and MMHH] and Eq.~\eqref{eq:weighted_Manhattan_average}~[MHHM]. The second metrics uses the known local error rates and calculates shortest weighted paths calculated with Dijkstra algorithm.}
\label{table:thresholds}
\end{table}

\subsubsection{Benchmarking locally-tailored codes}\label{sec:benchmarks}

We have numerically simulated and assessed the performances of the CSS code, the XXZZ code, and noise-tailored MHHM and MMHH codes using a modified version of the QECSIM Python package~\cite{qecsim}. First, we benchmark the performance of the error correcting codes using a matching decoder with input weights defined according to the Manhattan distance. For square-shaped codes~(CSS, XXZZ, MMHH), we use the MWPM algorithm with the standard Manhattan weights calculated as in Eq.~\eqref{eq:weight_manhattan}. The MHHM code performs optimally on a rectangular lattice due the structure of its logical operators, similarly to the XZZX code under biased noise. Hence, for decoding a syndrome of the MHHM code we use the weighted Manhattan distance metrics similar to Eq.~\eqref{eq:weighted_Manhattan},
\begin{equation}\label{eq:weighted_Manhattan_average}
    w_{i,j} = -\log \langle p_h \rangle l_x - \log \langle p_m \rangle  l_z,
\end{equation}
where $\langle p_h \rangle$ and $\langle p_m \rangle$ are the probabilities of high- and medium-rate errors averaged over all qubits in the code. 

\begin{figure}[t]
    \centering
    \includegraphics[width=0.90\columnwidth]{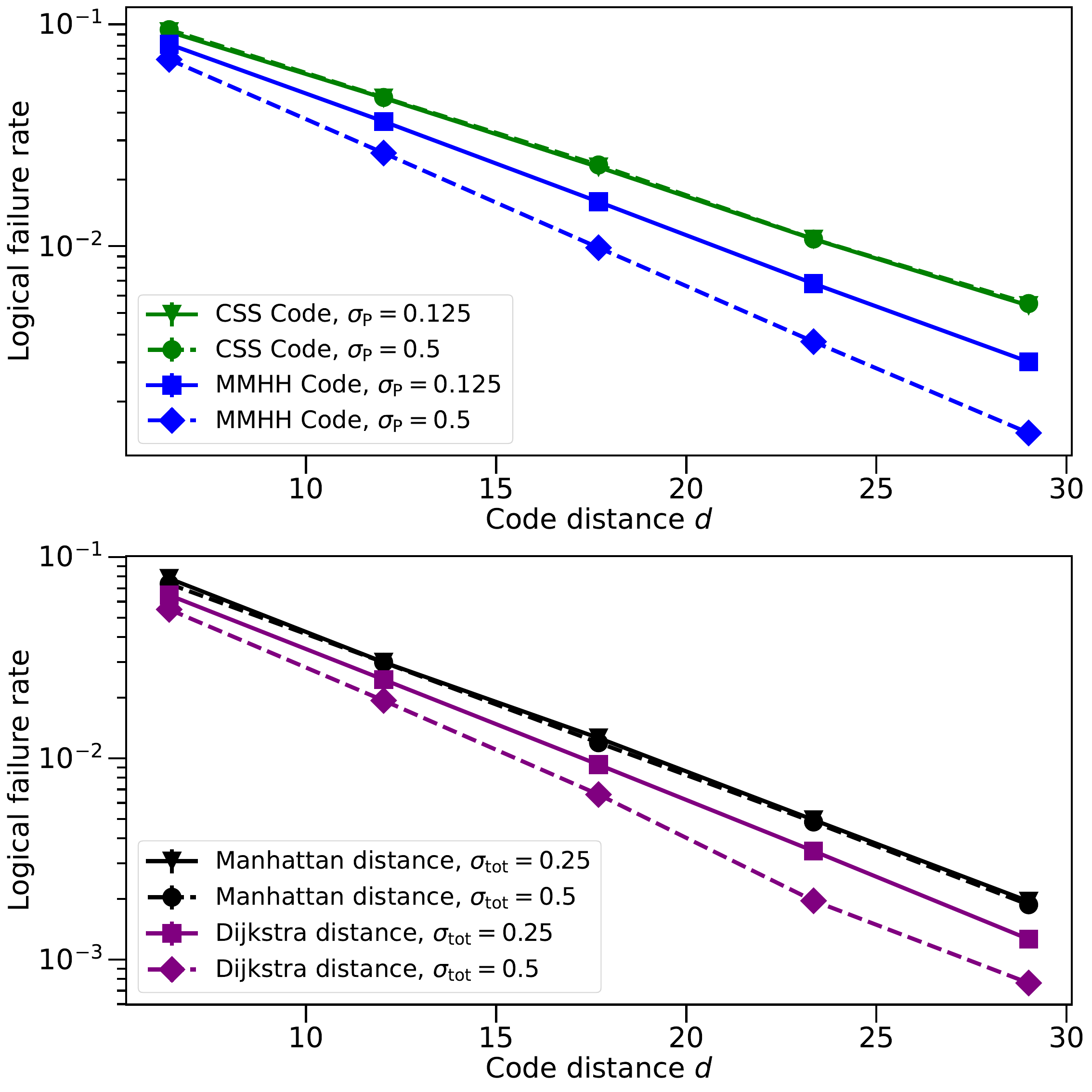}
    \caption{\label{fig:4} Sub-threshold logical failure rates 
    vs.~code distance $d$ of squared-shaped surface codes. (a)~Tailoring codes stabilizers to local Pauli noise. Green and blue curves correspond to, respectively, the CSS and noise-tailored MMHH codes. The curves are derived using a standard MWPM decoder with the Manhattan distance defined in Eq.~\eqref{eq:weight_manhattan}. Data is shown for $p = 0.1$ and $\sigma_{\textrm{tot}} = 1/2$ in Eq.~\eqref{eq:probabilities}. Solid and dashed lines correspond to $\sigma_{\textrm{P}} = 0.125$ and $\sigma_{\textrm{P}} = 0.5$, respectively. The XXZZ and XZZX codes yield logical error rates identical to those of the CSS code, while the MHHM code demonstrates a scaling similar to that of the MMHH code. These curves are hence not shown. (b)~Tailoring a matching decoder to local total noise. Black~(purple) curve correspond to the MHHM code in conjunction with a decoder using  
    the Manhattan~(Dijsktra) distance metrics. Solid and dashed lines correspond to noise models with $\sigma_{\textrm{tot}} = 0.25$ and $\sigma_{\textrm{tot}} = 0.5$, respectively.}
\end{figure}

Figure~\ref{fig:4}~(a) demonstrates the comparison between sub-threshold logical error rates of different variants of the surface code versus a code distance $d$. The noise-tailored codes clearly demonstrate exponentially suppressed logical failure rates when compared to the CSS and XXZZ codes. We also summarize the error correction thresholds for four descriptive sets of parameters $\sigma$ in Table~\ref{table:thresholds}. We observe that when the code stabilizers are tailored to non-uniform Pauli noise the thresholds are improved as well. As expected, the advantage of using the locally-tailored MHHM and MMHH codes over the CSS and XXZZ codes grows as we increase $\sigma_{\textrm{P}}$, both in terms of thresholds~[Fig.~\ref{fig:9}] and sub-threshold scaling~[Fig.~\ref{fig:4}~(a)]. 
On the contrary, changing $\sigma_{\textrm{tot}}$ does not affect the error correction capabilities of the codes; both the threshold and sub-threshold scaling remain intact as we vary $\sigma_{\textrm{tot}}$. In Appendix~\ref{sec:thresholds}, we explore the decoding capabilities of different codes for a wider range of parameters $\sigma$ and discuss the thresholds calculation procedure in more detail.

\subsection{Noise-aware matching decoder}\label{sec:tailored_decoder}

The improved performance of the surface code observed in Fig.~\ref{fig:4}~(a) and Appendix~\ref{sec:thresholds} is achieved solely by adapting the code stabilizers, while the decoder uses the distance metrics of Eqs.~\eqref{eq:weight_manhattan} and \eqref{eq:weighted_Manhattan_average}. That is, no information about local error rates has been provided to the decoder. However, a decoder can as well be tailored to more reliably decode error syndromes by utilizing the available information about the error model~\cite{PhysRevA.106.062428}. Here we investigate the behaviour of the surface codes in conjunction with a decoder that uses the local error rates of each qubit. The weights of the matching graph are calculated according to the probabilities of the shortest weighed paths using Dijkstra algorithm, which we implement with the PyMatching package~\cite{higgott2021pymatching}. 

Our numerical simulations show that implementing a noise-aware decoder instead of the one based on the Manhattan distance does not noticeably improve error correction capabilities of the code against non-uniform Pauli noise, that is, both threshold and sub-threshold scaling are nearly unaffected when we vary $\sigma_{\textrm{P}}$ at a fixed $\sigma_{\textrm{tot}}$. In contrast, such a modification yields improvement in the presence of non-uniform total errors, i.e., when we increase $\sigma_{\textrm{tot}}$. As such, Table~\ref{table:thresholds} demonstrates noticeably higher error thresholds enabled by a noise-aware decoder compared to the case of Manhattan distance for $\sigma_{\textrm{tot}}=0.5$. As we show in Fig.~\ref{fig:4}~(b), the sub-threshold error rates also scale more favourably when Dijkstra metrics is used, with logical error rate decreasing as $\sigma_{\textrm{tot}}$ grows. We explain this behaviour by the fact that errors become less random when $\sigma_{\textrm{tot}}$ is increased. As an illustrative example, consider a situation where the total error probabilities of certain qubits are close to zero. In such a case, the decoder knows that errors should not take place on these qubits and makes a more informative decision about locations of errors, resulting in more accurate decoding. For a detailed comparison of the two decoders, we refer the reader to Appendix~\ref{sec:Dijkstra} and figures therein. 

\subsection{Correction of non-uniform errors}\label{sec:inid}

Summarizing this Section, we considered two types of non-uniformity in qubit noise: variation in relative Pauli error rates and variation in total error rates of qubits. We have shown that the former can be countered by applying noise-aware Clifford deformations to the code stabilizers, and the latter can be corrected more efficiently by making the decoder noise-aware. The advantage of tailoring the code stabilizers become more apparent as we introduce stronger variations in Pauli noise~(increase $\sigma_{\textrm{P}}$), while using the noise-aware decoder becomes more advantageous as we introduce stronger variations in total noise~(increase $\sigma_{\textrm{tot}}$). Importantly, the modifications to the code stabilizers and the decoder input weights are independent and can hence be implemented simultaneously to correct arbitrary non-identically distributed noise. Further steps can be made along the same lines of research. As such, the residual correlations that take place between the dual and primal sub-lattices due to low-rate Pauli errors can be partially taken into account by an advanced version of the MWPM decoder~\cite{6874997}. Furthermore, local Clifford deformations can in principle be combined with belief propagation or belief matching~\cite{higgott2022fragile}, which might lead to an improved accuracy, but are likely to be much more computationally demanding than the fast matching decoder used in our simulations. We leave the question of designing alternative decoding algorithms outside of the scope of this paper. Finally, we note that here we have focused on investigating the properties of the surface code on a non-rotated lattice, as in Fig.~\ref{fig:1}. The sub-threshold scaling can be further improved by applying analogous local Clifford deformations on the surface code with rotated geometry. In particular, it has been shown that the number of physical qubits required to achieved a given logical error rate is reduced by half when the rotated code configuration is used~\cite{PhysRevA.76.012305}.

\section{Correlated errors and code degeneracy}\label{sec:correlated}

We now turn to the case where correlated errors are present between data qubits of the surface code. For concreteness, we focus on two-qubit interactions between nearest-neighbour qubits. Such two-qubit noise is architecture specific and occurs due to various physical mechanisms. In superconducting chips, unwanted cross-talk correlations can take place when a CNOT gate is executed via cross-resonance~(CR) technique~\cite{PhysRevLett.107.080502,PhysRevB.81.134507}. Qubits involved in the CR gate can not be efficiently isolated from their environment and are inevitably exposed to neighbouring qubits via a $ZZ$ interaction~\cite{10.1145/3503222.3507761,PhysRevApplied.12.054023}. Particularly, in the context of surface-code error correction, a CNOT gate between data and ancilla qubits can cause unwanted $ZZ$ error between the data qubit involved in the gate and one of the neighbouring data qubits~\cite{PhysRevLett.127.060505}. Similar cross-talk noise is ubiquitous in other hardware platforms as well. In silicon spin qubits, $ZZ$ cross-talk errors can occur due to microwave-induced phase shifts~\cite{Xue2022}. In a trapped ion system, laser intensity spillover onto the neighbouring ions during gate applications can lead to unwanted $XX$-type cross-talk errors between the qubits involved in the desired gate and their neighbours~\cite{Grzesiak2020}. 

The effect of correlated noise on surface-code error correction has been analysed in a very recent experimental work~\cite{harper2023learning}. The authors have benchmarked correlations that occur between data qubits during the operation of a 39-qubit Google Sycamore quantum processor and concluded the crucial importance of taking nearest-neighbour correlated errors into account for accurate estimations of the performance of error correction protocols. On the other hand, correlated errors on qubits located further apart are less probable and expected to have less destructive effect on the error correction fidelity, as discussed in Ref.~\cite{PhysRevA.89.042334}. Hence, our focus here is on local two-qubit errors only. 

Consider qubits subject to a combination of single-qubit depolarizing noise
\begin{equation}\label{eq:single_q_depol}
    \mathcal{E}_m^{(1)}(\rho)
    =
    \frac{1}{3}
    \sum_{k=X,Y,Z}
    \sigma_m^{k}
    \rho
    \sigma_m^{k}
\end{equation}
and two-qubit noise of the form
\begin{equation}
\label{eq:2_qubit_XX_ZZ}
\begin{aligned}
    \mathcal{E}_m^{(2)}(\rho)
    &=
    p_{XX}
    \sum_{l \in \mathcal{N}_m}
    X_m
    X_{l}
    \rho
    X_m
    X_{l}
    \\&+
    p_{ZZ}
    \sum_{l \in \mathcal{N}_m}
    Z_m
    Z_{l}
    \rho
    Z_m
    Z_{l}.
    \end{aligned}
\end{equation}
Here, $\mathcal{N}_m\subset L$ is a set of all 
nearest neighbors of $m\in L$ and $p_{XX}, p_{ZZ} \in[0,1]$ are the probabilities of correlated two-qubit bit- and phase-flip errors, respectively. Our choice of two-qubit noise model~\eqref{eq:2_qubit_XX_ZZ} is motivated by experimental data, showing that $XX$ and $ZZ$ are the most ubiquitous types of two-qubit errors omnipresent in various hardware platforms, as discussed above. The total noise channel affecting qubit $m$ then reads
\begin{equation}\label{eq:two_qubit}
    \begin{aligned}
    \mathcal{E}(\rho)
    &=
    (1-p_1 - p_2)
    \rho
    \\&+
    p_1\mathcal{E}_m^{(1)}(\rho)
    +
    p_2\mathcal{E}_m^{(2)}(\rho),
    \end{aligned}
\end{equation}
where $p_1$ and $p_2 = p_{XX} + p_{ZZ}$ are the probabilities of single- and two-qubit errors, respectively. 

In a weakly interacting regime, $p_2/p_1 \ll 1$, the dynamics are dominated by single-qubit errors and the probability of connecting two defects via the most probable chain of errors reads
\begin{equation}\label{eq:P1}
P_1
=
\begin{pmatrix}
l_x + l_z
\\
l_x
\end{pmatrix}
p_1^{l_x + l_z}.
\end{equation}
This expression is identical to Eq.~\eqref{eq:prob_css} up to a pre-factor$ \begin{pmatrix}
l_x + l_z
\\
l_x
\end{pmatrix}$, which takes into account the number of shortest strings of single-qubit errors connecting two stabilizers and will be referred to as the \emph{degeneracy factor}. As explained in Ref.~\cite{Criger2018}---and confirmed by our simulations---accounting for degeneracy does not have effect on correcting single-qubit errors in the surface code with the standard geometry. Hence the degeneracy factor was omitted in the previous sections, leading to the Manhattan distance of the form~\eqref{eq:weight_manhattan}. On the other hand, below we will show that accounting for degeneracy is crucial for efficient correction of correlated errors, or when correction is performed on the rotated surface-code lattice. 

\begin{figure}
    \centering
    \includegraphics[width=0.95\columnwidth]{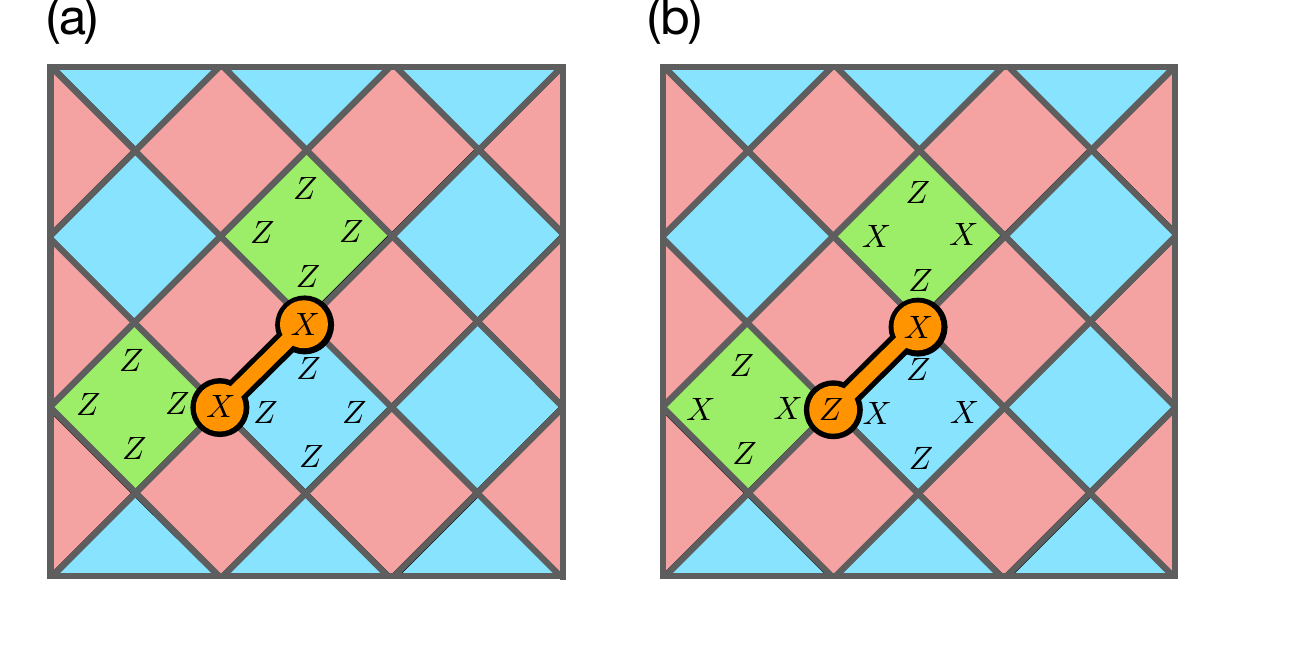}
    \caption{\label{fig:5} Surface codes tailored for a known two-qubit noise channel. (a)~Two-qubit phase-flip and bit-flip errors between nearest neighbors create anyons within one of the sub-lattices, as shown with green tiles, and can be efficiently decoded. (b)~In the presence of, e.g., XZ correlations, the code stabilizers can be Clifford-deformed to maintain the same decoding capabilities as in (a).}
\end{figure}

Assume now a regime of strong interactions, where $p_2/p_1 \gg 1$. From Fig.~\ref{fig:5}~(a), it is straightforward to check that the noise described by the quantum channel~\eqref{eq:2_qubit_XX_ZZ} flips dual and primal stabilizers of the CSS code in pairs, producing a syndrome well suitable for the MWPM decoder. The probability of connecting the defects within each sub-lattice  via a chain of two-qubit errors can be deduced from Fig.~\ref{fig:5}~(a) and reads
\begin{equation}\label{eq:P2}
P_2
=
\begin{pmatrix}
l_m
\\
(l_x + l_z)/2
\end{pmatrix}
p_2
^{l_m},
\end{equation}
where $l_m = \mathrm{max}(l_x,l_z)$ and the pre-factor accounts for the degeneracy of shortest strings composed of two-qubit errors. The input weights of the MPWM can then be approximated by taking into account single- and two-qubit error stings and read
\begin{equation}\label{eq:weights_mod}
w_{i,j}^{(\textrm{deg+corr)}}
=
-\log(P_1 + P_2),
\end{equation}
which is valid to the lowest order in the limits of weak~($p_2/p_1 \ll 1$) and strong~($p_2/p_1 \gg 1$) coupling regimes~\cite{note_first_order}.

\begin{figure}[t]
    \centering
    \includegraphics[width=0.95\columnwidth]{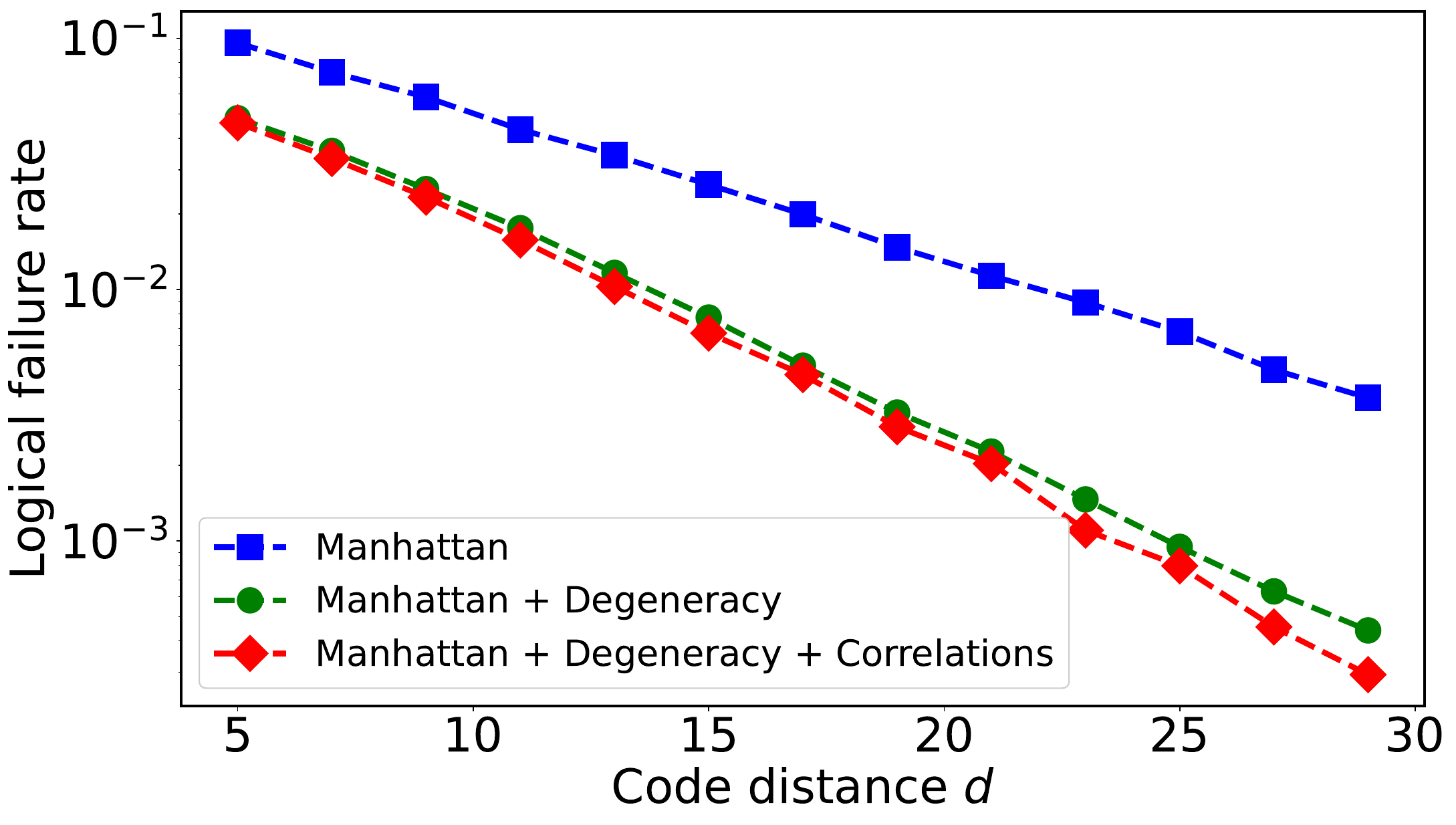}
    \caption{\label{fig:6} Simulated logical error rates versus code distance in the presence of nearest-neighbor correlations. Qubits are affected by a combination of single-qubit depolarizing noise and local two-qubit bit- and phase-flip errors, as described by Eq.~\eqref{eq:two_qubit}. When used in conjunction with the standard CSS code, such two-qubit noise produces stabilizers defects in pairs, making the syndrome well suitable for the MWPM decoder. The blue~(squares) curve corresponds to the MWPM with edges weighted according to the standard Manhattan distance~\eqref{eq:weight_manhattan}. The green~(circles) curve corresponds to the weights that take syndrome degeneracy into account, i.e., defined according to Eq.~\eqref{eq:weights_deg}. The red~(diamonds) curve corresponds to a decoder that explicitly takes into account the probability of two-qubit errors, as in Eq.~\eqref{eq:weights_mod}. Data is shown for $p = 0.125$ and $p_1 = 0.25p$.}
\end{figure}

We numerically investigate the performance of the CSS surface code in the presence of noise described by a quantum channel of Eqs.~\eqref{eq:two_qubit} and \eqref{eq:2_qubit_XX_ZZ} with $p_{XX} = p_{YY} = 0.5$. We benchmark the performance of the surface code accompanied by a matching decoder that uses (i)~the standard Manhattan distance metrics of Eq.~\eqref{eq:weight_manhattan} and (ii)~distance determined according to Eq.~\eqref{eq:weights_mod}, i.e., that takes single- and two- qubit errors with the corresponding degeneracies into account. As expected, the latter decoder yields much more efficient error correction capabilities when two-qubit errors are present, see Fig.~\ref{fig:6}. This result is in agreement with the observation of Ref.~\cite{PhysRevA.89.042334}.

Our simulations, however, show a surprising feature of error correction in the presence of correlated two-qubit noise: the term $P_2$ in Eq.~\eqref{eq:weights_mod} is, in fact, redundant, and ignoring it has vanishingly small effect on the performance of the decoder, as shown in Fig.~\ref{fig:6}. Instead, merely accounting for syndrome degeneracy in Eq.~\eqref{eq:P1} is responsible for the suppressed logical error rates. Ignoring the probability of connecting two defects by a chain of two-qubit errors $P_2$, Eq.~\eqref{eq:weights_mod} reduces to
\begin{equation}\label{eq:weights_deg}
    w_{i,j}^{\textrm{deg}} 
    =
    l_x+l_y - \log{
    \begin{pmatrix}
    l_x + l_y
    \\
    l_x
    \end{pmatrix}
    },
\end{equation}
where we have omitted a constant shift $-\log p_1>0$. As we show in Fig.~\ref{fig:6}, matching decoders with weights defined according to Eqs.~\eqref{eq:weights_mod} and~\eqref{eq:weights_deg} perform nearly identically, meaning that no information about the magnitudes of single- and two-qubit noise is required by the decoder to efficiently decode a syndrome in the presence of correlated two-qubit noise. 

\begin{figure}[t]
    \centering
    \includegraphics[width=0.95\columnwidth]{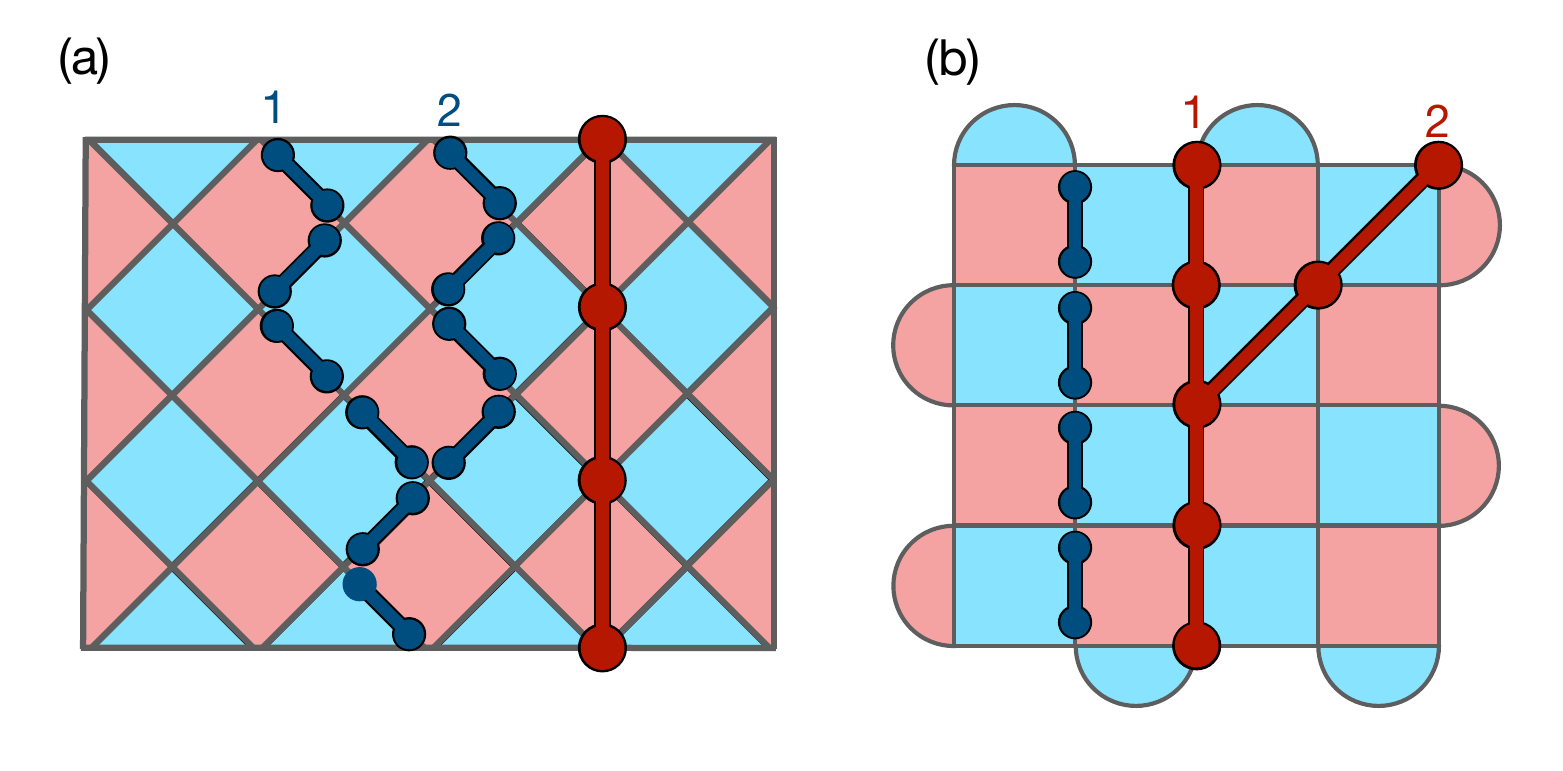}
    \caption{\label{fig:7} Examples of logical $\lx$ operators of the (a)~non-rotated and (b)~rotated surface codes. Single- and two-qubit errors are shown schematically with red dots and blue dumbbells, respectively. In (a), single-qubit error can form only $d$ different shortest-path logical operators, while the number of shortest-path logical operators formed by two-qubit errors is highly degenerate. For the rotated surface code of (b), the situation is reversed, i.e., the number of logical errors due to single-qubit errors is highly degenerate.}
\end{figure}

To explain this peculiar feature, we first turn to the case of single-qubit noise present in two configurations of the surface code: the standard~[Fig.~\ref{fig:7}~(a)] and the rotated~[Fig.~\ref{fig:7}~(b)] surface codes. The effect of the syndrome degeneracy in two such configurations has recently been investigated in Ref.~\cite{Criger2018}. The authors have shown that taking syndrome degeneracy into account improves the decoding accuracy for the case of the rotated surface code, but has no effect for the standard code, which agrees well with our numerical simulations provided in Appendix~\ref{sec:rotated_vs_nonrotated}. The difference in effect that syndrome degeneracy has on decoding can be attributed to degeneracy of logical operators produced by single-qubit errors. Figure~\ref{fig:7} shows examples of such logical operators realized by shortest chains of single-qubit errors in the two configurations. While the standard surface code only supports $d$ such logical errors, the rotated code can support a much larger number of them due to a high degree of logical operator degeneracy. Intuitively, in the former case a minimum-length path of single-qubit errors that traverses at least $d/2$ qubit locations occurs with much higher probability due to degeneracy. Hence, to adequately assess the error probabilities, degeneracy has to be taken into account in the case of the rotated code.

A similar, but opposite effect takes place when correlated errors are present in the system: Local two-qubit errors form highly-degenerate shortest-path logical operators in the non-rotated code, while the rotated code only supports $d$ shortest two-qubit chains, see Fig.~\ref{fig:7}. Hence, taking the degeneracy into account in Eq.~\eqref{eq:weights_deg} has a positive effect of decoding two-qubit errors in the non-rotated surface code, which agrees with the results of Fig.~\ref{fig:6}. Following the same logic, one can expect that taking degeneracy into account for decoding similar correlated errors in the rotated code will have negligible effect of the accuracy. This also agrees well with our numerical simulations provided in Appendix~\ref{sec:rotated_vs_nonrotated}. Hence, taking code degeneracy into account plays a similar and important role for decoding single-qubit error on the rotated code as for decoding two-qubit error on the non-rotated code. Put another way, turning on correlated errors can be thought of as partial rotation of the code in the context of logical operator degeneracy. To summarize, Ref.~\cite{Criger2018} has shown that the choice of boundary conditions influences decoder performance. Here, we generalize this observation by showing that the performance is determined by a combination of the error model and the boundary conditions. 

As shown in Fig.~\ref{fig:5}~(a), the standard CSS surface code used in conjunction with a MPWM decoder is well suited for correcting two-qubit noise described by a quantum channel~\eqref{eq:2_qubit_XX_ZZ}. It is, however, not unreasonable to assume an experimental setup, or an algorithm, where the leading two-qubit error takes the form of an $XZ$ channel, e.g., due to design aspects of a particular hardware implementation, or as a hook error copied from an ancilla qubit during stabilizer measurements~\cite{Trout_2018}. Such a two-qubit error of XZ-type described by a quantum channel
\begin{equation}\label{eq:2_qubit_XZ}
\begin{aligned}
    \mathcal{E}_m^{(2)}(\rho)
    &=
    \frac{1}{2}
    \sum_{l \in \mathcal{N}_m}
    \Big{(}
    X_m
    Z_{l}
    \rho
    Z_l
    X_{m}
    \\&+
    Z_m
    X_{l}
    \rho
    X_l
    Z_{m}
    \Big{)}
\end{aligned}
\end{equation}
creates four defects in both dual and primal sub-lattices of the CSS code, hence compromising the performance of a matching decoder. To maintain the compatibility with a decoder, the code can be Clifford-deformed. For example, as we show in Fig.~\ref{fig:5}~(b), the XZZX obeys the same dynamics under XZ correlations as the CSS code under two-qubit bit-flip and phase-flip channel~\eqref{eq:2_qubit_XX_ZZ}. Hence, knowing the type of the leading two-qubit noise, the surface code can be made compatible with the MPWM decoder under such noise by choosing an appropriate Clifford deformation.

As we have shown above, a MWPM decoder can efficiently handle correlated nearest-neighbor errors of certain types without the need of adjusting the decoding procedure to a known structure of the noise. Instead, merely taking syndrome degeneracy into account effectively adjusts the matching decoder to a noise model. Physical qubits in the surface code can in principle be affected by correlated errors other than two-qubit Pauli errors. In the future, it will be interesting to investigate if a similar argument holds when qubits are affected by correlated errors of different structures.

\section{Conclusion}\label{sec:conclusions}

In conclusion, we have studied the performance of Clifford-deformed surface codes in the presence of non-identically distributed and correlated two-qubit errors. In both cases the sub-threshold logical error rates are shown to be exponentially suppressed compared to the standard codes. Hence, the code footprint, i.e., the number of physical qubits required to achieve a target logical error rate, is exponentially smaller than in the case of the CSS code. Importantly, the improved decoding accuracy is achieved using a fast matching decoder that does not require numerically extensive pre-processing and can potentially be implemented in real quantum hardware. We have studied the performance of matching decoders with input parameters defined according to different distance metrics and shown the advantage of using the shortest weighted paths as graph weights in the case of non-uniform qubit error rates. Our aim has been to demonstrate the principle of noise-tailored local Clifford deformations. To this aim, we have used simple error models as test cases. This can be seen as a further substantial step towards designing quantum error correction schemes optimized for a particular hardware implementation. Finally, we have investigated the connection between code degeneracy and structures of the noise affecting qubits in the rotated and non-rotated surface code configurations. This in turn allowed us to identify cases where taking code degeneracy into account has a positive effect on decoding accuracy. 

In this work, we have considered the performance of noise-tailored surface codes for simple \textit{code capacity} noise models where only the data qubits are subject to error. As a direction for further work, it would be interesting to investigate noise-tailored codes under more realistic circuit level noise models for which it is assumed every location in the error correction circuit has the potential to introduce errors. At the circuit level, an important design consideration is the choice of two-qubit entangling gates. If an entangling operation turns one species of Pauli errors to another, it can trigger error propagation events that cause the noise channel to become unbiased. This can diminish the benefits of bias-tailoring. Fortunately, \textit{bias-preserving} hardware implementations of entangling gates have been proposed that are designed to maintain the system's noise-asymmetry. For instance, it is shown in Ref.~\cite{Puri2020} that bias-preserving gates can be achieved using stabilized cat qubits in driven nonlinear oscillators. This architecture has been demonstrated to integrate well with the XZZX code~\cite{PRXQuantum.2.030345}, and it would be interesting to investigate similar schemes in the setting of the Clifford-deformed codes explored in this work.    

The notion of Clifford deformations makes sense also for other \emph{topological quantum error correcting codes} beyond the toric or surface codes. For example, various instances of Clifford-deformed 3D codes have been studied in \cite{huang2022tailoring}. Bias-tailoring has also been shown to be a useful technique for quantum low density parity check codes \cite{roffe2023biastailored}. Many of the developed ideas of Clifford-deformed codes beyond the iid setting readily carry over to other codes as well, giving rise to new sets of  commuting stabilizers that are adapted to the noise model. It also seems perfectly conceivable that \emph{coherent errors} can be addressed in a way similar to the one described here. Additionally, it would be interesting to investigate the influence of Clifford deformations on the performance of other decoders such as the union find decoder and belief matching. 
It is the hope that the present work can contribute to the line of thought of bringing notions of quantum error correction closer to ideas on actual quantum hardware development, and to think about quantum error correction in more physically motivated terms.

\begin{acknowledgments}
We gratefully acknowledge funding by the Federal Ministry of Education and Research (BMBF) via the RealistiQ (13N15580), for which this is a joint work between the two project 
partners. J.~E.~has also been funded by the BMBF under QSolid and MUNIQC-ATOMS as well as by the DFG (CRC 183). The project is also part of the Munich Quantum Valley (K8), which is supported by the Bavarian state government with funds from the Hightech Agenda Bayern Plus. The authors thank the HPC Service of ZEDAT, Freie Universit{\"a}t Berlin, for offering computing time~\cite{Bennett2020}.
\end{acknowledgments}

\section*{Code and data availability}
The code used for numerical simulations of different QECCs discussed in this work is available at~\cite{simulations_code}. The data and the scripts for generating the plots are available at~\cite{plotting_code}.

\bibliographystyle{apsrev4-2}
\bibliography{citations}

\begin{thebibliography}{81}
\bibitem [{not({\natexlab{a}})}]{note_coherent_noise}%
  \BibitemOpen
  \bibinfo {note} {For coherent noise, one could also consider more general
  Clifford conjugations, either by other unitaries from $C_1/U(1)$, or by
  conjugating several qubits at once and considering $C_n/U(1)$ for $n\geq 1$.
  Such code deformations will not be considered here.}\BibitemShut {Stop}%
\bibitem [{not({\natexlab{c}})}]{note_first_order}%
  \BibitemOpen
  \bibinfo {note} {In Eq.~\eqref{eq:weights_mod}, we only include the zeroth
  order terms in $p_1$ and $p_2$. In Ref.~\cite{PhysRevA.89.042334}, the
  probability of connecting two defects by a chain of single- and two-qubit
  errors has been calculated to the the higher order. That is, the authors have
  also included the possibility of creating connecting two defects with
  Manhattan distance $N$ by one single-qubit error and $N-1$ two-qubit errors
  when $p_1/p_2 \ll 1$~(by one two-qubit error and $N-1$ single-qubit errors
  when $p_2/p_1 \ll 1$). However, our simulations show that adding such
  higher-order terms has vahishingly small effect on the decoding
  fidelity.}\BibitemShut {Stop}%
\end{thebibliography}

\onecolumn
\newpage
\appendix

\section{Error thresholds calculation}\label{sec:thresholds}

We obtain the threshold estimates of Table~\ref{table:thresholds} using the critical exponent method described in Refs.~\cite{WANG200331,harrington2004analysis}. Specifically, for physical error rates near the threshold we fit the logical error rate to the 
function $P_{\textrm{fit}}: \mathbb{R}^+\rightarrow \mathbb{R}$ with
\begin{equation}
    P_{\textrm{fit}} (x) = B_0 + B_1 x + B_2 x^2,
\end{equation}
where
\begin{equation}
x := (p - p_{\textrm{th}})^{\alpha d ^{\beta}}.
\end{equation}
Here, $d$ is the code distance, $p$ is the single-qubit error rate, and $B_0,B_1,B_2,\alpha,\beta$ are fitting parameters.

\begin{figure}[h!]
    \centering
    \includegraphics[width=0.9\textwidth]{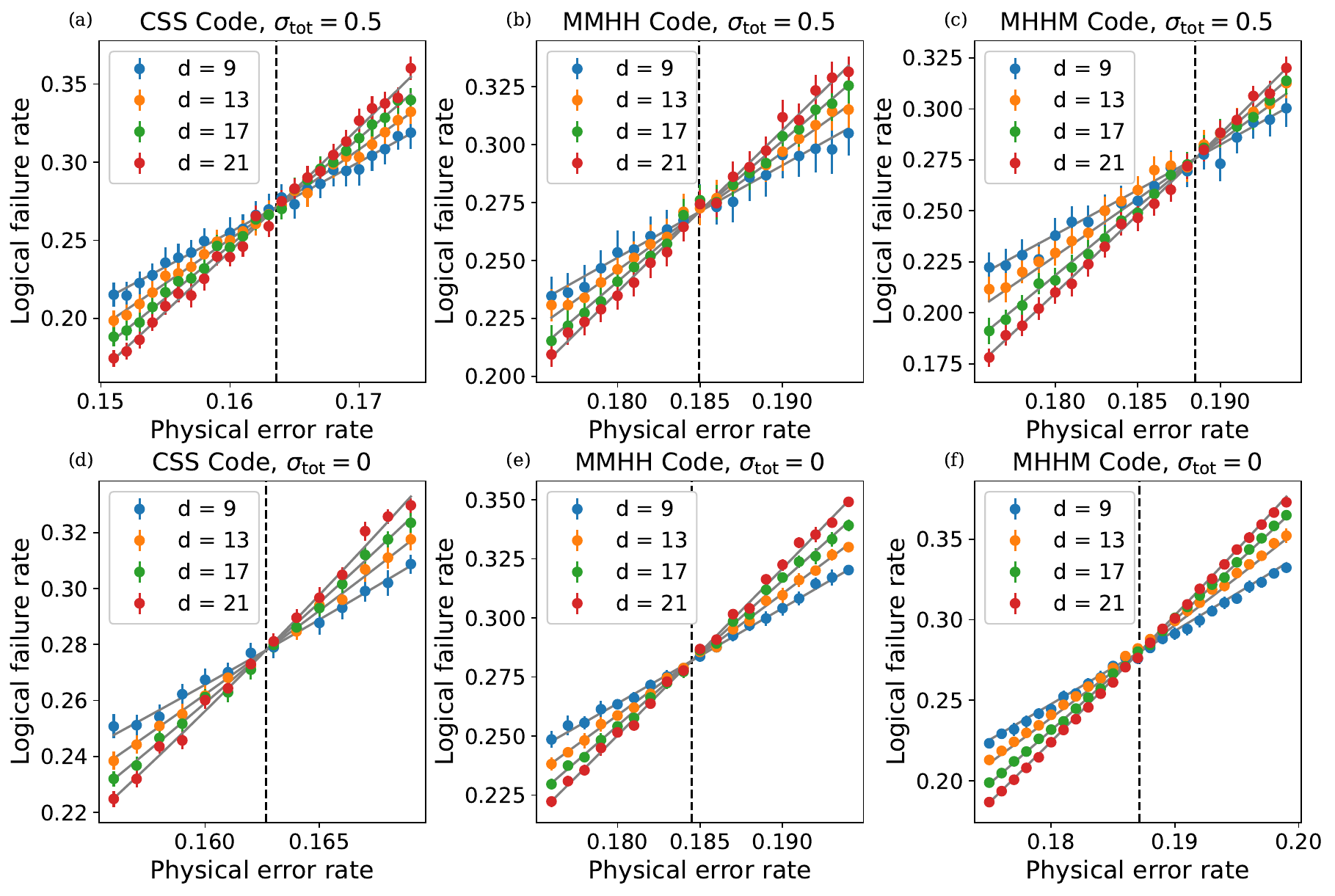}
    \caption{\label{fig:8} Logical error rates of the (a,d)~CSS and (b,e)~MMHH, and (c,f)~MHHM surface codes in the presence of non-identically distributed Pauli errors with $\sigma_{\textrm{P}}=0.5$. In panels (a)--(c) the total qubit noise is nonuniform with $\sigma_{\textrm{tot}}=0.5$. In panels (d)--(f) the total qubit noise is uniform. The syndromes are decoded using MWPM with distance metric calculated according to the Manhattan distance. Error bars show one standard deviation.}
\end{figure}

\begin{figure}[h!]
    \centering
    \includegraphics[width=0.65\textwidth]{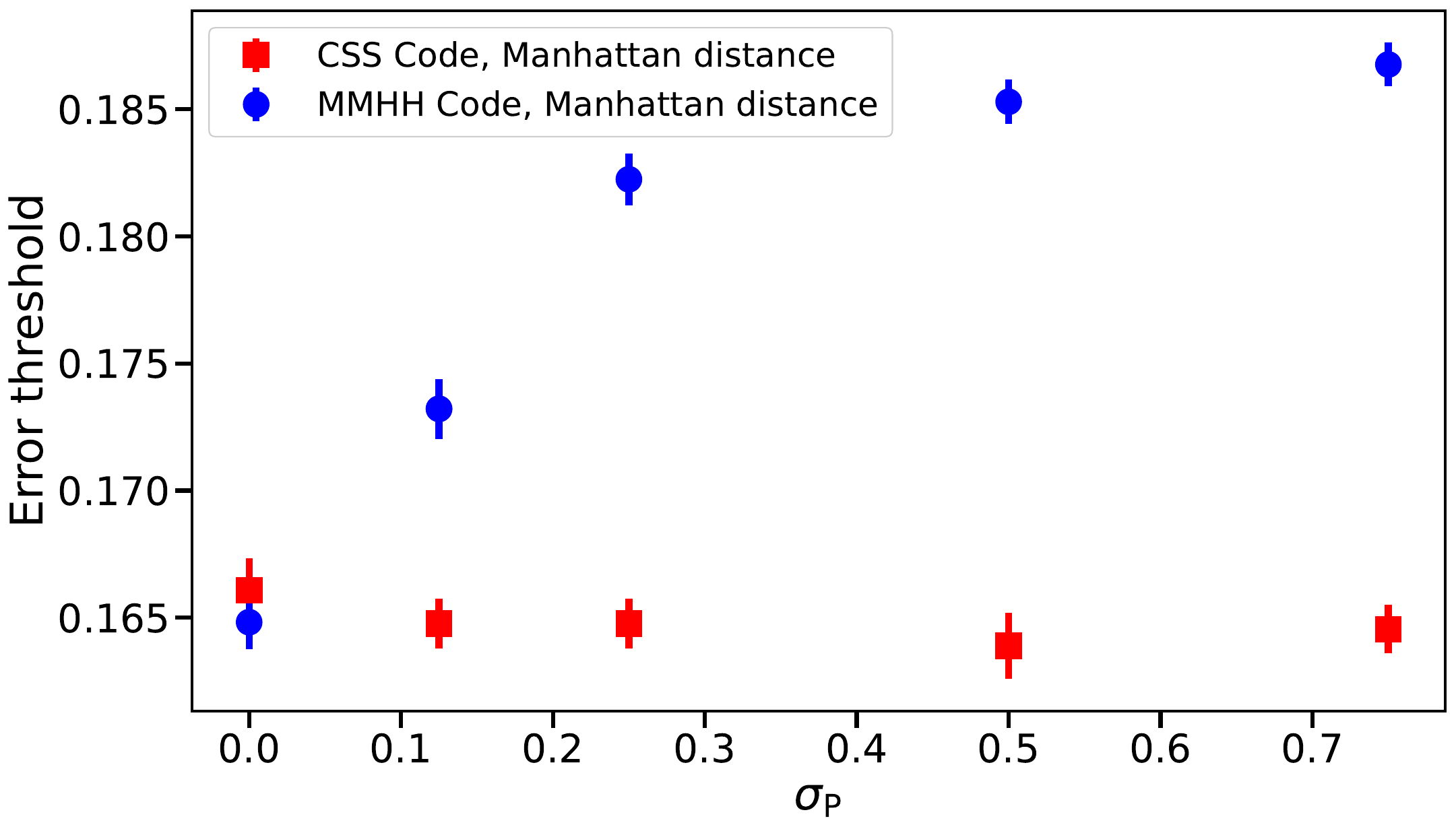}
    \caption{\label{fig:9} Error correction thresholds of the CSS~(red squares) and MMHH~(blue circles) surface codes versus the standard deviation $\sigma_{\textrm{P}}$ in Eq.~\eqref{eq:probabilities} with $\sigma_{\textrm{tot}}=1/2$. Thresholds are calculated using a MWPM matching decoder with input weights defined according to the Manhattan distance. The threshold of the MMHH increases with the variation in relative Pauli noise described by $\sigma_{\textrm{P}}$, while the threshold of the CSS code remains constant.}
\end{figure}

Figure~\ref{fig:8} shows near-threshold logical failure rates and threshold estimates of different codes. The thresholds are also summarized in Table~\ref{table:thresholds}. We see that the noise-tailored surface codes demonstrate error correcting thresholds exceeding those of the CSS code by $\approx 2.3\%$ for $\sigma_{\textrm{P}}=0.5$. In Fig.~\ref{fig:9} we compare the thresholds of the codes for a wider range of values $\sigma_{\textrm{P}}$. We observe increasing thresholds of the noise-tailored codes as $\sigma_{\textrm{P}}$ grows. On the other hand, we see little to no difference between error thresholds derived for non-uniform~($\sigma_{\textrm{tot}}=0.5$) and uniform~($\sigma_{\textrm{tot}}=0$) total error rates, meaning that Clifford deformations do not help with correcting variations in total error rates. 

\clearpage
\newpage

\section{Shortest-path matching}\label{sec:Dijkstra}

\begin{figure}[h!]
    \centering
    \includegraphics[width=0.95\textwidth]{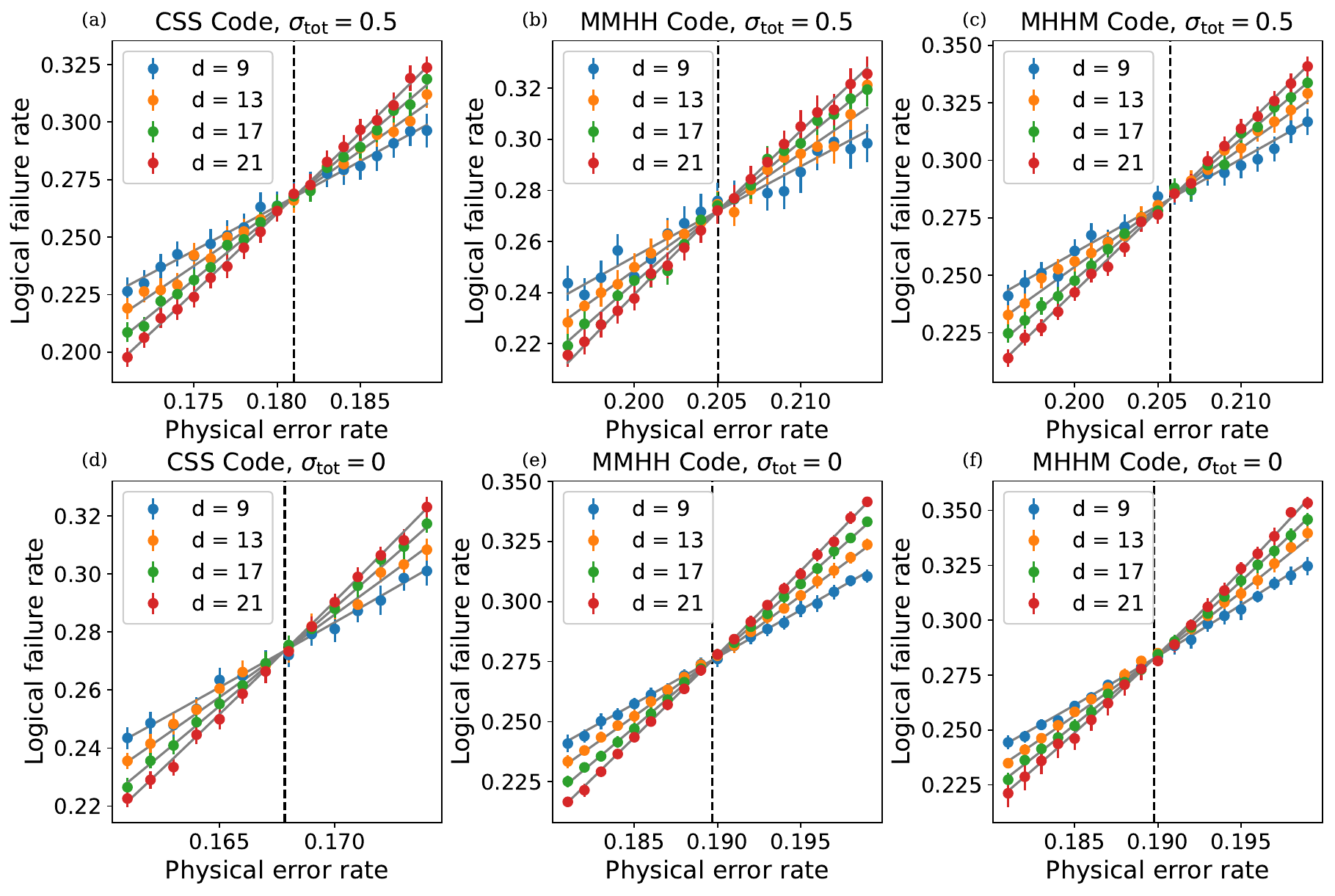}
    \caption{\label{fig:10} Same as Fig.~\ref{fig:8}, but for a matching decoder that uses input weights defined according to the shortest weighted paths. The parameters of the noise used for simulations are identical to those used in Fig.~\ref{fig:8}.}
\end{figure}

\begin{figure}[h!]
    \centering
    \includegraphics[width=0.5\textwidth]{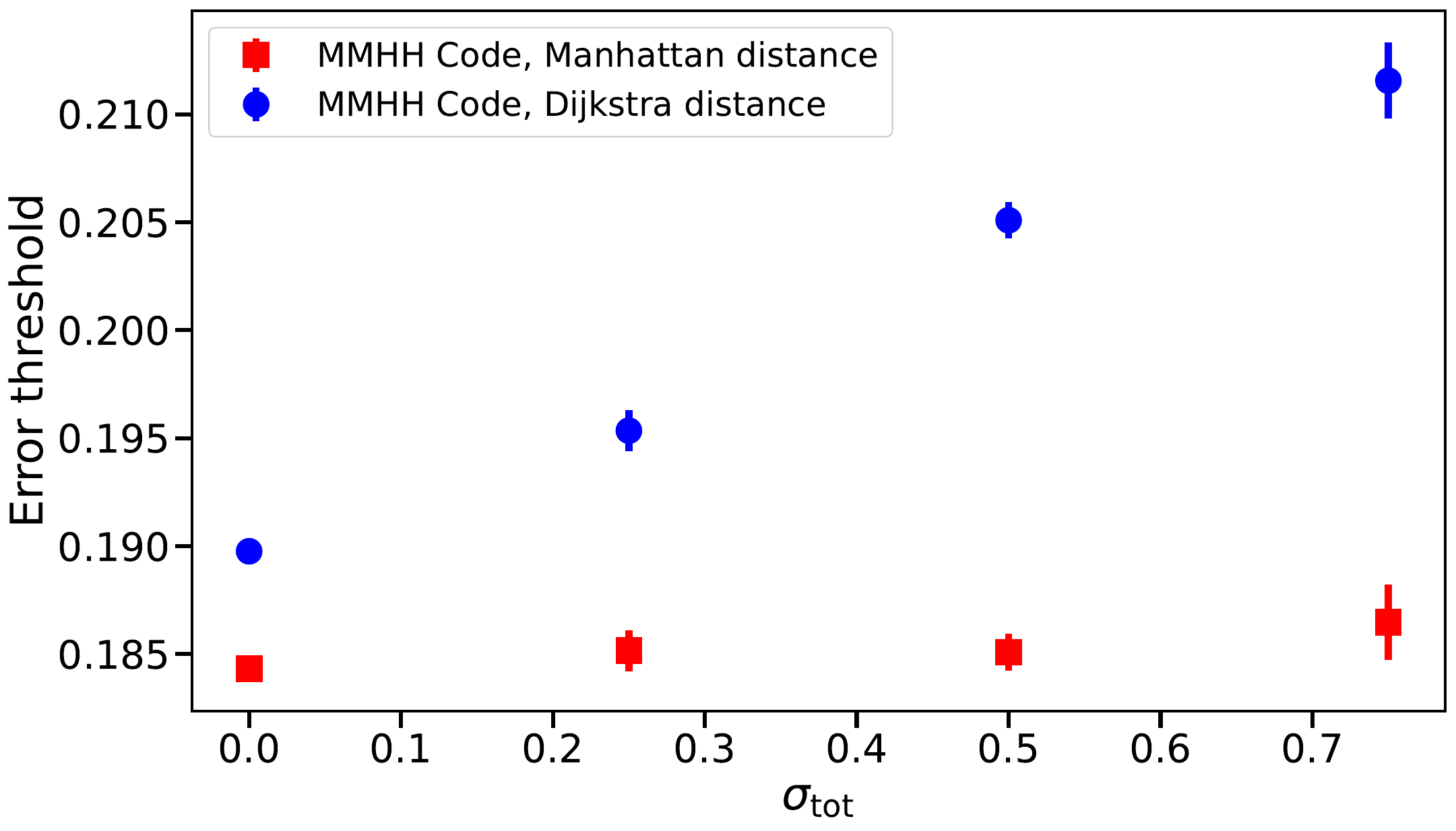}
    \caption{\label{fig:11} Error correction thresholds of the MMHH code versus the standard deviation of the total error rate $\sigma_{\textrm{tot}}$ in Eq.~\eqref{eq:probabilities_var} with $\sigma_{\textrm{P}}$. Red squares and blue circles correspond to matching decoders that use input weights defined according to the Manhattan and Dijkstra distance metrics, respectively. Other codes demonstrate qualitatively similar behaviour and are not shown.}
\end{figure}

Figure~\ref{fig:10} provides estimation of the error-correction threshold of different codes in the case when we use a noise-aware decoder and define the input weights according to the Dijkstra distance. In contrast to the decoder that uses Manhattan distance metrics, the decoder with Dijkstra-based metrics demonstrates a clear improvement of the error correction thresholds as $\sigma_{\textrm{tot}}$ grows, see Figs.~\ref{fig:10} and ~\ref{fig:11}. 

\clearpage
\newpage

\section{Code geometry and syndrome degeneracy}\label{sec:rotated_vs_nonrotated}

\begin{figure}[h!]
    \centering
    \includegraphics[width=0.75\textwidth]{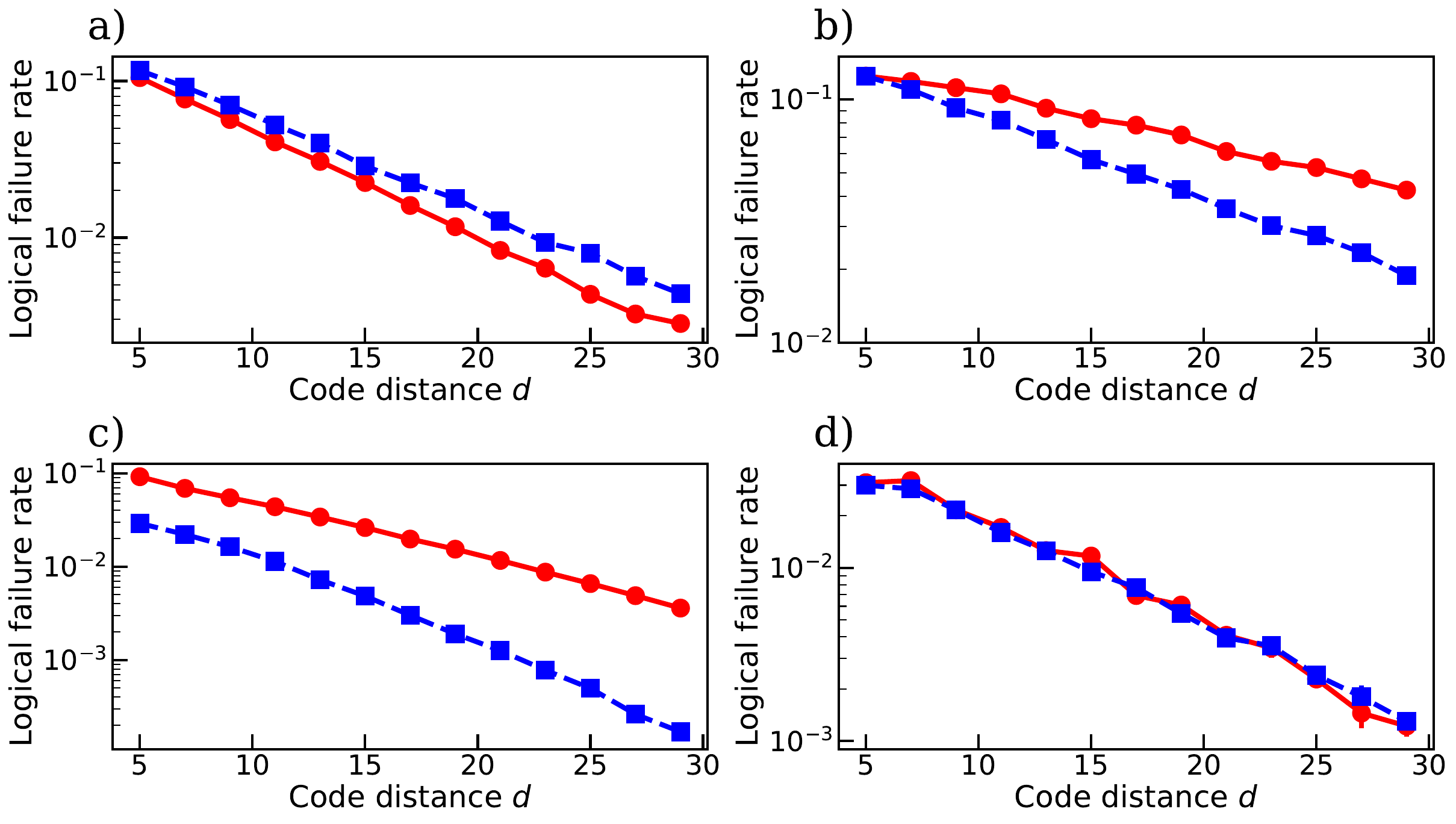}
\caption{\label{fig:12} Numerically simulated logical failure rates of the surface code with and without syndrome degeneracy taken into account. Different panels correspond to (a,c)~non-rotated and (b,d)~rotated surface codes in the presence of (a,b)~single-qubit and (c,d)~two-qubit Pauli errors. In the presence of single-qubit errors, adding the degeneracy term has little, negative effect on the decoding fidelity when the non-rotated code~(a) is used, while noticeably enhances it in the rotated code~(b). When the code is subject to two-qubit errors, adding the degeneracy  term has an opposite effect, as shown in (c,d). Here, code stabilizers are chosen according to Fig.~\ref{fig:5}, such that only one of the sub-lattices is affected by a single two-qubit error event. To generate the data in panels (a,b,c) the physical error rate is set to $p = 0.1$  and in (d) $p=0.05$.}
\end{figure}

In Fig.~\ref{fig:12}, we investigate the impact of 
code geometry and syndrome degeneracy. One can observe a qualitatively opposite behaviour when degeneracy is accounted for decoding error syndromes in rotated and non-rotated surface codes. 

\clearpage
\newpage

\begin{section}{Carbon footprint of numerical simulations}\label{sec:footprint}
The carbon footprint associated with the numerical
simulations in this paper are summarised below.
\begin{center}
\resizebox{!}{!}{%
\begin{tabular}[b]{l c}
\hline
\textbf{Numerical simulations} & \\
\hline
Total Kernel Hours [$\mathrm{h}$]& 917323\\
Thermal Design Power Per Kernel [$\mathrm{W}$]& 5.75\\
Total Energy Consumption Simulations [$\mathrm{kWh}$] & 5275\\
Average Emission Of CO$_2$ in Germany [$\mathrm{kg/kWh}$]& 0.56\\
Were The Emissions Offset? & Yes\\
\hline
\hline
Total CO$_2$-Emission [$\mathrm{kg}$] & 2954 \\
\hline
\hline
\end{tabular}}
\end{center}

\noindent Guidance on the reporting the carbon cost of scientific research can be found in Ref.~\cite{CO2}.
\end{section}

\end{document}